\documentclass[twocolumn]{aastex701} 

\usepackage{hyperref}
\usepackage{enumitem}
\usepackage{amssymb}
\usepackage{lipsum}

\newcommand{\um}{$\mu$m}
\newcommand{\spherex}{SPHEREx}

\newcommand{\hto}{H$_2$O}
\def\h2{H$_2$}
\def\co2{CO$_2$}
\def\ch3oh{CH$_3$OH}
\def\nh3{NH$_3$}
\newdimen\saa  \newdimen\sbb
\def\arcsec{\ifmmode {^{\scriptstyle\prime\prime}}
          \else $^{\scriptstyle\prime\prime}$\fi}
\def\arcmin{\ifmmode {^{\scriptstyle\prime}}
          \else $^{\scriptstyle\prime}$\fi}               
\def\parcs{\saa=.07em \sbb=.03em
     \ifmmode \hbox{\rlap{.}}^{\scriptstyle\prime\kern -\sbb\prime}\hbox{\kern -\saa}
     \else \rlap{.}$^{\scriptstyle\prime\kern -\sbb\prime}$\kern -\saa\fi}
\def\deg{\ifmmode^\circ\else$^\circ$\fi}

\accepted{to the Astrophysical Journal March 11, 2026}

\begin{document}
\makebox[17cm][r]{\textcopyright 2025 all rights reserved.}

\title{\spherex\ Wide-Field Infrared Spectral Mapping of Interstellar Ices and Polycyclic Aromatic Hydrocarbons}


\author[0000-0002-5599-4650]{Joseph L. Hora}
\email{jhora@cfa.harvard.edu}
\affiliation{Center for Astrophysics $|$  Harvard {\&} Smithsonian, 60 Garden St., Cambridge, MA 02138-1516, USA}
\author[0009-0007-1206-9506]{Jinyoung K. Noh}
\affiliation{Department of Physics and Astronomy, Seoul National University, 1 Gwanak-ro, Gwanak-gu, Seoul 08826, Korea}%
\email{kyla2001@snu.ac.kr}
\author[0000-0002-6025-0680]{Gary J. Melnick}
\email{gmelnick@cfa.harvard.edu}
\affiliation{Center for Astrophysics $|$  Harvard {\&} Smithsonian, 60 Garden St., Cambridge, MA 02138-1516, USA}
\author[0000-0001-7449-4638]{Brandon S. Hensley}
\affiliation{Jet Propulsion Laboratory, California Institute of Technology, 4800 Oak Grove Drive, Pasadena, CA 91109, USA}%
\email{brandon.s.hensley@jpl.nasa.gov}
\author[0000-0002-5158-243X]{Roberta Paladini}
\email{paladini@ipac.caltech.edu}
\affiliation{IPAC, California Institute of Technology, MC 100-22, 1200 E California Blvd, Pasadena, CA 91125, USA}%
\author[0000-0003-3119-2087]{Jeong-Eun Lee}%
\affiliation{Department of Physics and Astronomy, Seoul National University, 1 Gwanak-ro, Gwanak-gu, Seoul 08826, Korea}%
\email{lee.jeongeun@snu.ac.kr}%

\author[0000-0002-3993-0745]{Matthew L. N. Ashby}
\email{mashby@cfa.harvard.edu}
\affiliation{Center for Astrophysics $|$  Harvard {\&} Smithsonian, 60 Garden St., Cambridge, MA 02138-1516, USA}
\author[0000-0003-1841-2241]{Volker Tolls}
\email{vtolls@cfa.harvard.edu}
\affiliation{Center for Astrophysics $|$  Harvard {\&} Smithsonian, 60 Garden St., Cambridge, MA 02138-1516, USA}
\author[0000-0001-8064-2801]{Jaeyeong Kim}
\email{jaeyeong.kim@cfa.harvard.edu}
\affiliation{Korea Astronomy and Space Science Institute (KASI), 776 Daedeok-daero, Yuseong-gu, Daejeon 34055, Republic of Korea}
\affiliation{Center for Astrophysics $|$  Harvard {\&} Smithsonian, 60 Garden St., Cambridge, MA 02138-1516, USA}
\author[0000-0003-4990-189X]{Michael~W.~Werner}%
\affiliation{Jet Propulsion Laboratory, California Institute of Technology, 4800 Oak Grove Drive, Pasadena, CA 91109, USA}%
\email{michael.w.werner@jpl.nasa.gov}
\author[0000-0002-5710-5212]{James~J.~Bock}%
\affiliation{Department of Physics, California Institute of Technology, 1200 E. California Boulevard, Pasadena, CA 91125, USA}%
\affiliation{Jet Propulsion Laboratory, California Institute of Technology, 4800 Oak Grove Drive, Pasadena, CA 91109, USA}%
\email{jjb@astro.caltech.edu}%
\author[0000-0002-6503-5218]{Sean~Bruton}%
\affiliation{Department of Physics, California Institute of Technology, 1200 E. California Boulevard, Pasadena, CA 91125, USA}%
\email{sbruton@caltech.edu}%
\author[0009-0000-3415-2203]{Shuang-Shuang~Chen}
\affiliation{Department of Physics, California Institute of Technology, 1200 E. California Boulevard, Pasadena, CA 91125, USA}
\email{schen6@caltech.edu}%
\author[0000-0001-5929-4187]{Tzu-Ching~Chang}%
\affiliation{Jet Propulsion Laboratory, California Institute of Technology, 4800 Oak Grove Drive, Pasadena, CA 91109, USA}%
\affiliation{Department of Physics, California Institute of Technology, 1200 E. California Boulevard, Pasadena, CA 91125, USA}%
\email{tzu@caltech.edu}%
\author[0000-0001-6320-261X]{Yi-Kuan~Chiang}%
\affiliation{Academia Sinica Institute of Astronomy and Astrophysics (ASIAA), No. 1, Section 4, Roosevelt Road, Taipei 10617, Taiwan}%
\email{ykchiang@asiaa.sinica.edu.tw}%
\author[0000-0002-3892-0190]{Asantha~Cooray}%
\affiliation{Department of Physics \& Astronomy, University of California Irvine, Irvine CA 92697, USA}%
\email{acooray@uci.edu}
\author[0000-0002-4650-8518]{Brendan~P.~Crill}%
\affiliation{Jet Propulsion Laboratory, California Institute of Technology, 4800 Oak Grove Drive, Pasadena, CA 91109, USA}%
\affiliation{Department of Physics, California Institute of Technology, 1200 E. California Boulevard, Pasadena, CA 91125, USA}%
\email{bcrill@jpl.nasa.gov}%
\author[0000-0002-7471-719X]{Ari~J.~Cukierman}%
\affiliation{Department of Physics, California Institute of Technology, 1200 E. California Boulevard, Pasadena, CA 91125, USA}%
\email{ajcukier@caltech.edu}%
\author[0000-0001-7432-2932]{Olivier~Dor\'{e}}%
\affiliation{Jet Propulsion Laboratory, California Institute of Technology, 4800 Oak Grove Drive, Pasadena, CA 91109, USA}%
\affiliation{Department of Physics, California Institute of Technology, 1200 E. California Boulevard, Pasadena, CA 91125, USA}%
\email{olivier.dore@caltech.edu }%
\author[0000-0002-9382-9832]{Andreas~L.~Faisst}%
\affiliation{IPAC, California Institute of Technology, MC 100-22, 1200 E California Blvd, Pasadena, CA 91125, USA}%
\email{afaisst@caltech.edu}%
\author[0009-0009-1219-5128]{Zhaoyu~Huai}%
\affiliation{Department of Physics, California Institute of Technology, 1200 E. California Boulevard, Pasadena, CA 91125, USA}%
\email{zhuai@caltech.edu}%
\author[0000-0001-5812-1903]{Howard~Hui}%
\affiliation{Department of Physics, California Institute of Technology, 1200 E. California Boulevard, Pasadena, CA 91125, USA}%
\email{hhui@caltech.edu}%
\author[0000-0002-2770-808X]{Woong-Seob~Jeong}%
\affiliation{Korea Astronomy and Space Science Institute (KASI), 776 Daedeok-daero, Yuseong-gu, Daejeon 34055, Republic of Korea}%
\email{jeongws@kasi.re.kr}%
\author[0000-0002-5016-050X]{Miju~Kang}%
\affiliation{Korea Astronomy and Space Science Institute (KASI), 776 Daedeok-daero, Yuseong-gu, Daejeon 34055, Republic of Korea}%
\email{mjkang@kasi.re.kr}%
\author[0009-0003-8869-3651]{Phil~M.~Korngut}%
\affiliation{Department of Physics, California Institute of Technology, 1200 E. California Boulevard, Pasadena, CA 91125, USA}%
\email{pkorngut@caltech.edu}%
\author[0000-0002-3808-7143]{Ho-Gyu~Lee}%
\affiliation{Korea Astronomy and Space Science Institute (KASI), 776 Daedeok-daero, Yuseong-gu, Daejeon 34055, Republic of Korea}%
\email{hglee@kasi.re.kr}%
\author[0000-0002-9548-1526]{Carey~M.~Lisse}%
\affiliation{Johns Hopkins University, 3400 N Charles St, Baltimore, MD 21218, USA}%
\affiliation{Johns Hopkins University Applied Physics Laboratory, Laurel, MD 20723, USA}%
\email{carey.lisse@jhuapl.edu}%
\author[0000-0001-5382-6138]{Daniel~C.~Masters}%
\affiliation{IPAC, California Institute of Technology, MC 100-22, 1200 E California Blvd, Pasadena, CA 91125, USA}%
\email{dmasters@ipac.caltech.edu}%
\author[0009-0002-0149-9328]{Giulia~Murgia}%
\affiliation{Department of Physics, California Institute of Technology, 1200 E. California Boulevard, Pasadena, CA 91125, USA}%
\email{gmurgia@caltech.edu}%
\author[0000-0001-9368-3186]{Chi~H.~Nguyen}%
\affiliation{Department of Physics, California Institute of Technology, 1200 E. California Boulevard, Pasadena, CA 91125, USA}%
\email{chnguyen@caltech.edu}%
\author[0000-0003-4408-0463]{Zafar~Rustamkulov}%
\affiliation{IPAC, California Institute of Technology, MC 100-22, 1200 E California Blvd, Pasadena, CA 91125, USA}%
\email{zafar@caltech.edu}%
\author[0000-0002-0070-3246]{Ji~Yeon~Seok}%
\affiliation{Korea Astronomy and Space Science Institute (KASI), 776 Daedeok-daero, Yuseong-gu, Daejeon 34055, Republic of Korea}%
\email{jyseok@kasi.re.kr}%
\author[0000-0001-7254-1285]{Robin~Y.~Wen}%
\affiliation{Department of Physics, California Institute of Technology, 1200 E. California Boulevard, Pasadena, CA 91125, USA}%
\email{ywen@caltech.edu}%
\author[0000-0003-3078-2763]{Yujin~Yang}%
\affiliation{Korea Astronomy and Space Science Institute (KASI), 776 Daedeok-daero, Yuseong-gu, Daejeon 34055, Republic of Korea}%
\email{yyang@kasi.re.kr}%
\author[0000-0001-8253-1451]{Michael~Zemcov}%
\affiliation{School of Physics and Astronomy, Rochester Institute of Technology, 1 Lomb Memorial Dr., Rochester, NY 14623, USA}%
\affiliation{Jet Propulsion Laboratory, California Institute of Technology, 4800 Oak Grove Drive, Pasadena, CA 91109, USA}%
\email{mbzsps@rit.edu}
\begin{abstract}
We present some of the first infrared spectral maps acquired by SPHEREx.  These maps, which to our knowledge are the largest of their type ever compiled in the near-infrared, reveal multiple strong lines due to interstellar ices and polycyclic aromatic hydrocarbons (PAHs) throughout the Cygnus X and North American Nebula regions.  The maps emphasize the strongest features arising from the 3\,\micron\ \hto, 4.27\,\micron\ \co2, and 4.67\,\micron\ CO lines and the 3.28\,\micron\ PAH feature, all of which are detected over large areas with complex and filamentary spatial distributions. The ice absorption maps of \hto\ and \co2 in particular broadly trace dense, cold, and well-shielded regions across Cygnus X, consistent with the established picture of efficient ice formation in dense molecular clouds. The interstellar ice features are also detected abundantly in diffuse absorption over wide areas. The relative strength of the \hto\ and \co2\ features varies among different lines of sight, indicating possible differences in local physical conditions or chemical variations. The 3.28\,\micron\ PAH emission correlates with the emission from the 7.7 and 11.2\,\micron\ features, but shows small differences that may trace the grain size distribution and variations in the ambient UV field. SPHEREx all-sky spectral imaging, of which only a small fraction is showcased in this work, will support numerous science investigations including the structure of the Galaxy, the physics of the interstellar medium, and the chemistry of stars.   

\end{abstract}


\section{Introduction} 
\label{sec:intro}
\setcounter{footnote}{0}

Interstellar ices are thought to harbor significant amounts of prebiotic molecules—including water (\hto), carbon dioxide (\co2), carbon monoxide (CO), as well as smaller quantities of methanol (CH$_3$OH), methane (CH$_4$), ammonia (NH$_3$), cyanate (OCN$^-$), and carbonyl sulfide (OCS) \citep{2023McClure}. These ices, which coat dust grains participating in the formation of new stars and their surrounding protoplanetary disks, are considered the primary reservoirs of these key volatiles. The composition of interstellar ices – and consequently the inventory delivered to nascent planetary systems – is an important factor in understanding how planets such as our own acquire water and organic molecules.

These ice species exhibit characteristic broad absorption features which are diagnostic of their column densities and compositions. Because most of these key spectral features lie at wavelengths $>$2.5\,\micron, absorption in the Earth's atmosphere as well as high thermal backgrounds from the sky and telescopes make them difficult to observe from the ground. Therefore, many of the most sensitive existing measurements of ices in molecular clouds have come from infrared space missions, most notably the Infrared Space Observatory \citep[ISO; see][]{2004gibb} and Spitzer \citep[e.g.,][]{2013Boogert}. See the review by \citet{2015Boogert} for a summary of early ground- and space-based observations of ices. The composition of interstellar ices is a major focus of several JWST studies, which have already revealed significant variation in ice composition across different Galactic locations \citep[e.g.,][]{2024Bergner,2024Rocha}. 

\subsection{The SPHEREx Mission}

\spherex\footnote{\url{http://spherex.caltech.edu}}, the Spectro-Photometer for the History of the Universe, Epoch of Reionization, and ices Explorer \citep{2025Bock}, is a NASA mission optimized for and currently carrying out a groundbreaking all-sky spectroscopic survey from 0.75 to 5.0\,\um.  At the end of its two-year primary mission, \spherex\ will have mapped the sky four times to provide complete spectra of each 6\parcs{15} $\times$ 6\parcs{15} pixel on the sky.  The data returned by the mission will serve a variety of science objectives, including the main SPHEREx science themes: mapping the 3-dimensional structure of the  $z \lesssim 2$ universe to constrain the physics of inflation, performing intensity mapping to trace the development of star formation in galaxies over cosmic history, and tracing the evolution of water and other molecular ices from the interstellar medium (ISM) into star systems.

\begin{figure*}
    \centering
    \includegraphics[width=1.00\linewidth]{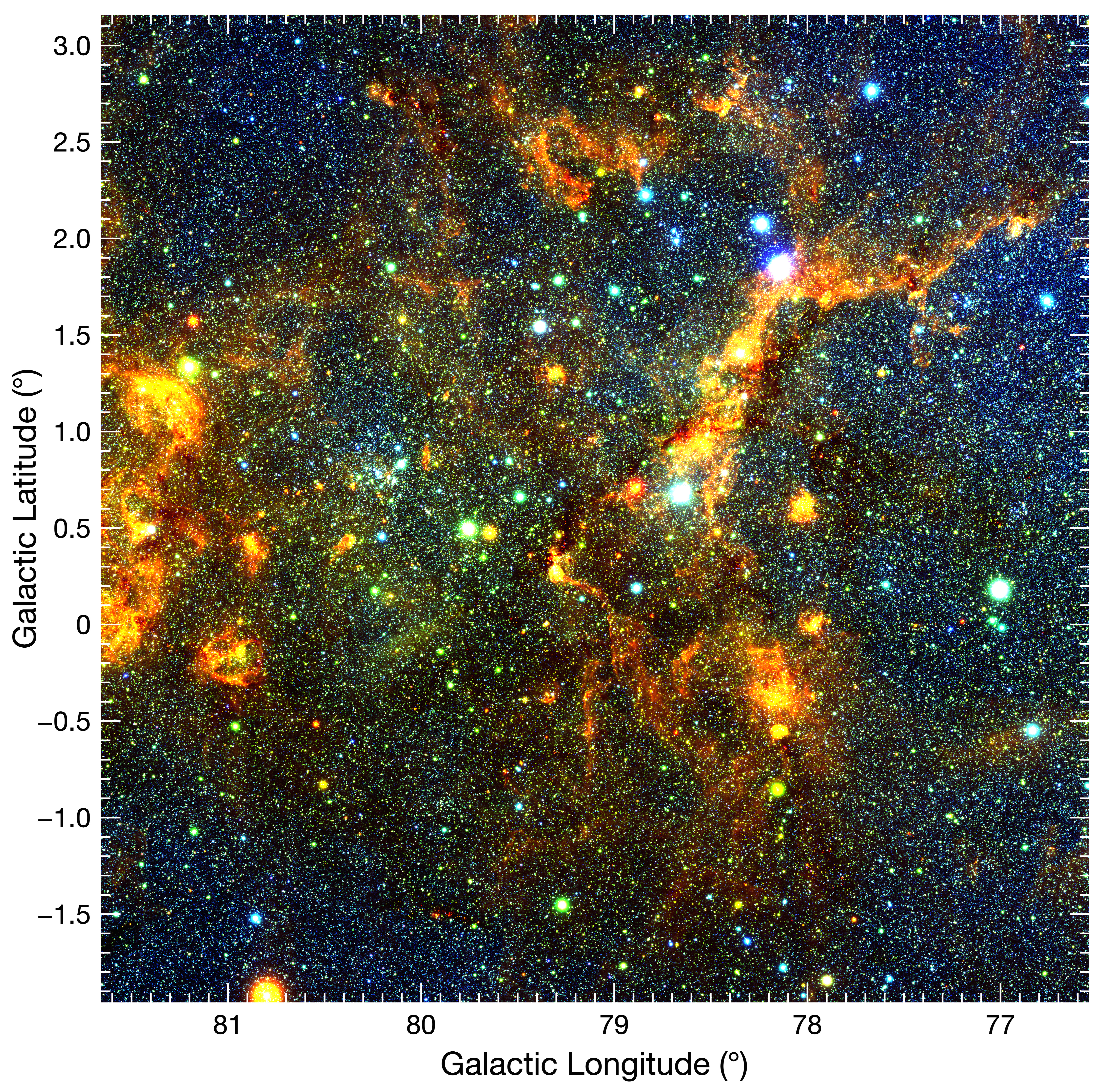}
    \caption{A three-color image of the CygX region constructed from SPHEREx mosaics. In this image, blue is  1.1 - 1.22\,\micron, green is  2.8 - 3.20 \,\micron\  (containing the \hto\ ice absorption feature), and red is 4.75 - 5.1\,\micron. Continuum emission from warm dust extends across the image as evidenced by the extended orange clouds. Dark lanes are seen where the dust and \hto\ ice absorption are the strongest. Red sources appear in these dark lanes where there are background sources or embedded young stars shining through the dust. Blue stars are seen in regions with lower extinction or are foreground objects.}
    \label{fig:colorcyg}
\end{figure*}

The SPHEREx Ices Investigation \citep{2026Melnick} will obtain ice absorption spectra along nearly 10 million preselected lines of sight. The targets were chosen based on evidence for intervening dust (inferred from their broadband colors), spatial isolation, and sufficient flux to ensure high signal-to-noise spectra \citep[][]{2023mashby}. This approach will generate a statistically significant dataset of ice spectra across a wide variety of Galactic environments, enabling the identification of key factors that influence the evolution of ice composition.

\subsection{Wide-Field Mapping of Spectral Features}
With its full-sky survey capability, SPHEREx is particularly well-suited to complement JWST's high-resolution, small-area observations by delivering large-area coverage. The NIRSpec integral field unit (IFU) mode covers a similar 1 - 5\,\micron\ range with a maximum resolution of $R=2700$ compared to SPHEREx's $R=130$, but its field of view (FoV) is 3$\times$3 arcsec, approximately 1/4 the area of a SPHEREx pixel. The SPHEREx FoV covers an area of the sky over 50 million times larger in each exposure.

While the aforementioned strategy of studying ices through their absorption imprints on spectra of isolated background sources remains the most accurate method of extracting column densities, initial viewing of the SPHEREx data made obvious that there is much more signal to be harvested from this novel type of dataset.  
From a quick glance at the first SPHEREx images of the Galactic plane it was obvious that SPHEREx would be a powerful tool for mapping the distribution of interstellar ices and polycyclic aromatic hydrocarbon (PAH) emission, as shown when the first images from SPHEREx were presented by \citet{2025Korngut}. (Note: in this paper we refer to the 3.28\,\micron\ emission as the PAH feature as it is commonly referred to in the literature, although the identity of the carrier(s) of this feature is still under some debate; e.g., see \citealt{2025Tokunaga}). 
Even in the raw images one can see dark clouds in the Galactic plane due to the broad 3\,\micron\ \hto\ feature. Filaments and bubbles in the ISM light up in the 3.28\,\micron\ PAH feature (including aliphatic hydrocarbons) as the SPHEREx FoV sweeps across the Galaxy. By combining the SPHEREx images into wide-field mosaics, we can trace the distribution of these spectral features over large areas in the Galactic plane.

As shown below (\S\ref{sec:mosaics}), SPHEREx is able to construct ice absorption maps on large spatial scales, 
a unique capability offered by no other facility.  Most earlier studies of ice features rely on observations of discrete continuum sources behind molecular clouds and in the envelopes and disks of young stellar objects (YSOs) to detect the absorption bands \citep[e.g., see the review by][]{2015Boogert}. Limited mapping has been performed by using multiple obscured sources, for example in Taurus \citep{2000Murakawa} and in Ophiuchus \citep{2006Pontoppidan}. 
\citet{2013Noble} constructed column density maps of \hto, \co2, and CO ices in several molecular cores using the spectroscopy mode of AKARI. In addition, \citet{2008Sonnentrucker} used Spitzer/IRS to construct a fully-sampled ice map in a 50\arcsec$\times$50\arcsec\ field within the star-forming region Cepheus A East using ice absorption features in the 5 -- 20\,\micron\ range.

Several notable ice maps have recently been constructed using JWST.  Recently, \citet{2025Smith} observed multiple lines of sight towards the Chamaeleon\,I molecular cloud and mapped \hto, \co2, and CO ice absorption in a $\sim$2\arcmin$\times$2\arcmin\ area. Also, studies of dark clouds in the Galactic center, including The Brick \citep{2023Ginsburg,2025Ginsburg} and the star-forming filament G0.342+0.024 \citep{2025Gramze} demonstrated that ice column densities can be derived from JWST photometry, mapping regions up to 8\arcmin\ across. \citet{2025Gunay} have proposed that absorption maps (as well as maps of the 3.4\,\micron\ aliphatic hydrocarbon and 10\,\micron\ silicate features) could be constructed from NIRCam and MIRI imaging using their medium- and wide-band filter sets.

However, NIRCam and MIRI have FoVs of 129\arcsec$\times$129\arcsec\ and 73\arcsec$\times$113\arcsec, respectively, and must acquire multiple filter images for each line (to estimate the continuum and the line depth), so JWST mapping of degree-sized fields would be extremely time-intensive. The SPHEREx ice mosaics will complement the targeted Ices Investigation by constraining the large-scale spatial distributions of ice species and their relationship with one another,  providing critical context for understanding the environments in which these ices form and evolve.

\subsection{The Cygnus-X Region}\label{sec:CygX}
In this paper, we demonstrate some of the potential of SPHEREx for Galactic studies by focusing on Cygnus-X (hereafter CygX), a relatively nearby Milky Way star-forming region at a distance of $\sim$1.4\,kpc \citep{2012Rygl}. The SPHEREx launch date and survey plan \citep{2025Bryan} made CygX visible to SPHEREx early in the mission.

CygX is one of the most massive of the nearby ($<2$~kpc) star forming regions \citep[see review by][]{2008Reipurth}, and contains many \ion{H}{2} regions, hundreds of massive protostars \citep{2014Kryukova}, thousands of low-mass YSOs \citep{2010Beerer, 2021Kuhn}, and several OB associations, including the Cyg OB2 cluster which alone contains over 50 O-type stars and hundreds of B stars \citep{2010Wright,2015Wright}.  

Recently, \citet{2024Zhang} performed a comprehensive study of the physical properties and 3D structure of molecular clouds towards the CygX region using CO survey and Gaia DR3 data. They determined distances using CO velocities and Gaia distances of associated stars, and find multiple gas layers present in the line of sight towards CygX. In particular, they found that the foreground Cygnus Rift structures located at 0.7 - 1~kpc contains $\sim$25\% of the molecular mass of the observed structures. Their distances are in general agreement with those determined from dust extinction measurements by \citet{2019Zucker,2020Zucker}. Many of the dark clouds seen against the background of emission from more distant CygX sources are in the Cygnus Rift layer, as we describe in the sections below.

\subsection{LDN 935}
We also constructed a mosaic and performed aperture photometry of an area in the North American and Pelican (NAP) Nebula complex \citep[for a review see][]{2008Reipurth}, a region of active star formation located at a distance of $\sim$800\,pc \citep{2020Zucker}. The SPHEREx map we present here is centered near Galactic coordinates ($l= 84$\fdg8, $b=-1$\fdg2) on a part of the dark cloud LDN\,935, which forms the ``Gulf of Mexico" that gives the nebula its distinctive shape at optical wavelengths.  The NAP complex contains an \ion{H}{2} region W80 which is being ionized by a massive O3.6 star \citep{2005Comeron,2016Maiz}.

Recently, \citet{2021Kong} presented a high-resolution CO molecular line survey of the NAP complex and assembled a detailed three-dimensional model of its structure. Much of LDN\,935 is revealed as composed of structures associated with the complex, likely located on the near edge of the W80 bubble. However, some parts of the nebula are at different velocities and are likely behind the bubble and not contributing to the extinction of sources immediately behind its near side. 

\section{Observations and Data Reduction}
\label{sec:Obs}
\subsection{SPHEREx Mosaics}
For the work described here we used the ``level 2" (L2) images and ``level 3'' (L3) spectra (\S\ref{sec:spectra}) produced by the SPHEREx Science Data Center (SSDC) pipeline \citep{2025Akeson}. 
The SPHEREx L2 images are publicly available at the IRSA web site\footnote{\url{https://irsa.ipac.caltech.edu/Missions/spherex.html}}. The SPHEREx explanatory supplement \citep{2025Exp} provides an overview of the L2 image characteristics and describes the data reduction pipeline. The data used here are from the QR2 version of the pipeline, the same version currently available on the IRSA archive.

The observations used in the Cygnus mosaics were obtained in two sets during the periods 2025 April 26 - 2025 July 5 and 2025 October 10 - 2025 November 2. All SPHEREx integrations use an exposure time of 113.5826\,s. For the mosaics constructed with wavelength ranges based on the SPHEREx spectral resolution as described below, an average of 148 individual exposures were used. For the LDN\,935 region, the data were obtained during 2025 May 9 - 2025 July 5 and 2025 October 23 - 2025 November 18, and an average of 52 exposures were used in each wavelength range.

\subsubsection{Custom Mosaic Construction}

We developed a custom set of python scripts to produce mosaics in wavelength ranges that facilitated analysis of the spectral features observed in the objects in the region of study. For example, to construct continuum-subtracted maps of narrow emission lines, we defined ranges centered on these lines, along with line-free continuum ranges at shorter and longer wavelengths that are used to estimate the continuum at the wavelength of the emission line. Certain features, such as the 3\,\micron\ \hto\ ice absorption, are several spectral resolution elements wide, so broader ranges were used in those mosaics to increase coverage and redundancy in the final images, and spectral regions were selected further from the line center to sample the continuum.

The first script reads each SPHEREx L2 image and slices it into separate images based on the defined wavelength ranges and the wavelength coverage of each band. The script examines the wavelength of peak transmission for each pixel, and if it is within the range specified, it is included in the output image. After these image slices are produced, a second script produces the mosaics from the image slices at the user-selected center position, size, and pixel scale. The mosaicking script  reads in the image slices, reprojects them to the final world coordinate system (WCS), and averages the slices at each pixel location to produce the mosaic. In cases where there are 3 or more samples at a location, a sigma-rejection is performed to eliminate outliers, which are mainly due to cosmic rays, satellite tracks, non-responsive pixels, or other artifacts that might not have been flagged in the L2 images. At this early stage of the SPHEREx mission, most of the observed areas of the sky have single coverage with small overlapping regions at the edges of neighboring scans. Therefore many of the images presented in this paper have visible artifacts such as pixels with undefined values due to the above mentioned effects. After SPHEREx has completed four scans of the sky, outlier pixels and artifacts can more easily be found and eliminated from the mosaics.
\label{sec:mosaics}

An example of an image constructed from SPHEREx L2 data is the 3-color CygX image shown in Figure~\ref{fig:colorcyg}. This image was constructed from mosaics that included the wavelength ranges 1.1 - 1.22\,\micron\ (blue), 2.8 - 3.20\,\micron\ (green), and 4.75 - 5.1\,\micron\ (red). The blue and red images are dominated by continuum emission from the stars and dust in the region, and the green image contains part of the broad \hto\ absorption feature centered near 3.0\,\micron, as well as stellar and dust continuum emission. 
The green image therefore is a mix of continuum emission from stars and warm dust, which is absorbed in the dark clouds and causes the background sources and objects embedded in the dark clouds to appear red, since the shorter wavelength emission is absorbed by the dust and \hto\ ice.

\subsubsection{``Banding'' Artifacts and Correction}
\label{sec:banding}
We constructed the mosaics presented here using pixels in discrete ranges of central wavelengths chosen to be wide enough to avoid spatial gaps in the images. These ranges have approximately the same widths as the pixel response function based on their positions on the LVF. The spectral features are therefore sampled at various points along the pixel spectral response curve that in general is not centered on the feature. This generates an artifact in the mosaics that makes the absorption depth (or emission feature strength) appear to vary across the range of center wavelengths, leading to curved stripes across the mosaics. After SPHEREx has mapped the sky several times, this effect can be minimized by constructing mosaics with much narrower wavelength ranges to keep the feature near the peak of the pixel response function at every point in the map. However, the mosaics presented here are constructed with the first survey data so only a single pass over each position of the sky was obtained, and some of these banding artifacts are visible.

\label{sec:bandcorr}
We have attempted to minimize this effect in the  \hto\ and \co2 peak optical depth images by applying a correction to the line images before calculating the optical depth. We constructed average absorption line profiles from higher-resolution JWST NIRSpec spectra downloaded from the public archive for a sample of 12 sources with ice absorption. Simulations of SPHEREx observations of the average line profiles were performed using the Ices SPHEREx simulator \citep{2026Tolls} to determine the factor needed to correct the observed absorption line depth to the true depth, which is dependent on the SPHEREx resolution at the central wavelength of the line. In addition, the simulations were used to determine a function to convert absorption depths measured at wavelengths displaced from the line center, given the line profile and SPHEREx resolution. We constructed difference images from the individual L2 exposures, applied the correction function to each pixel, and made a mosaic from the corrected difference images. We then calculated the peak optical depth image from the corrected absorption mosaic. 

\begin{figure}
    \centering
    \includegraphics[width=0.95\linewidth]{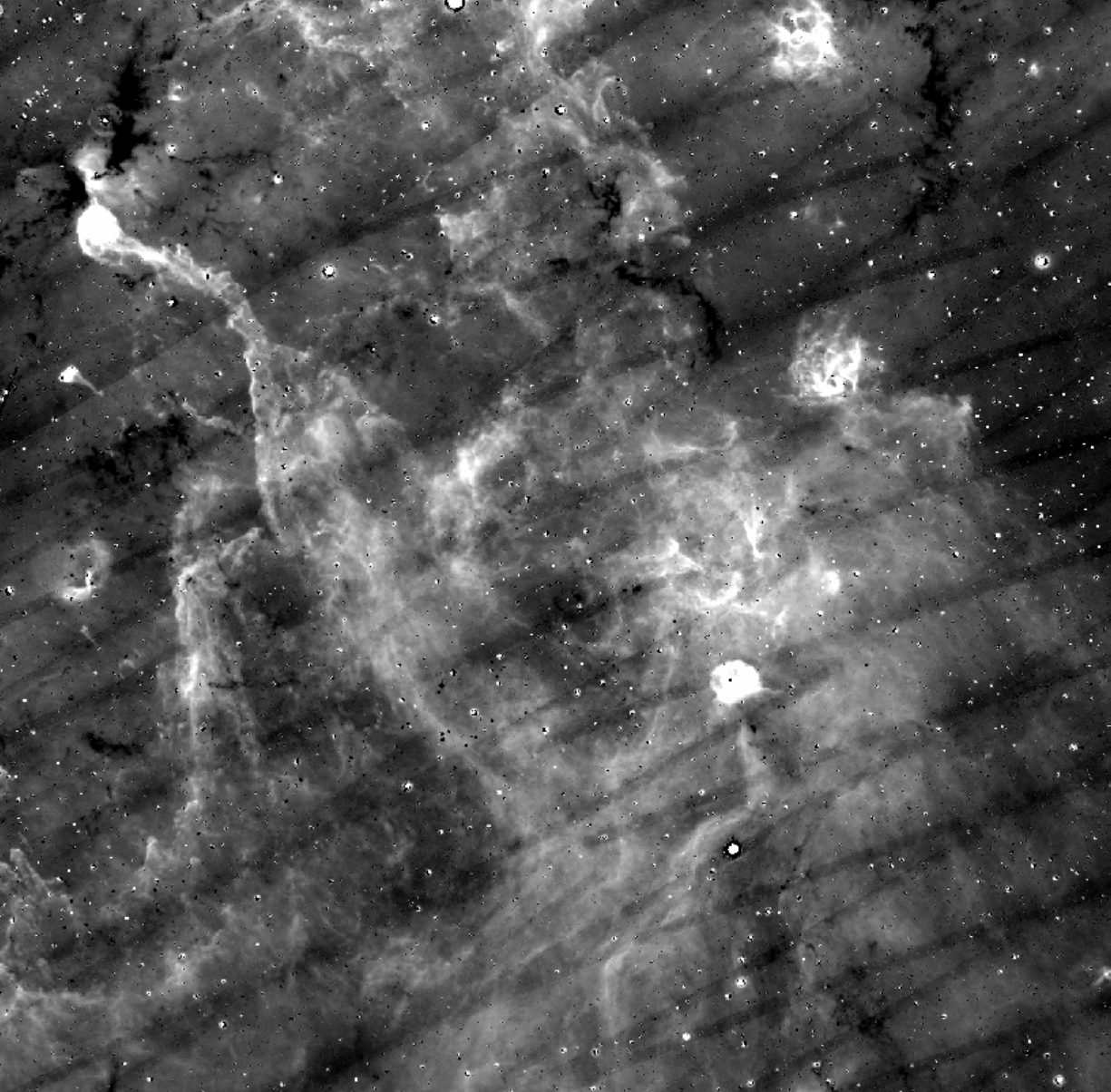}
    \includegraphics[width=0.95\linewidth]{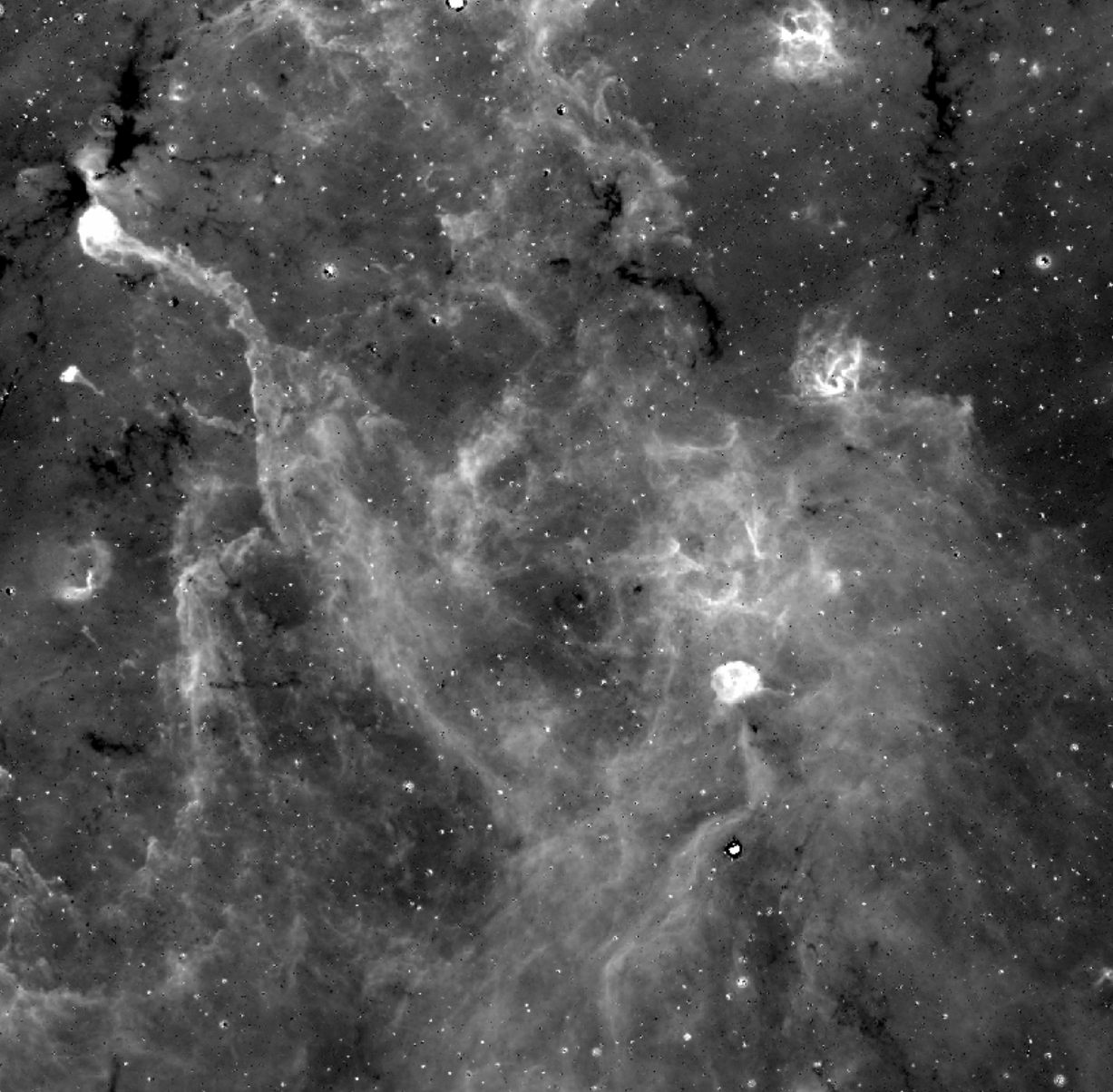}
    \caption{An illustration of the banding correction described in \S\ref{sec:bandcorr}. Top: an image of a 2\degr$\times$2\degr\ region near the center of the continuum-subtracted PAH mosaic, constructed using wavelengths in the range 3.24 - 3.34\,\micron. The dark bands appear at the wavelengths furthest from the peak of the PAH line. Bottom: the mosaic after applying the banding correction. }
    \label{fig:BandCorr}
\end{figure}

We used a similar process to produce the corrected PAH feature mosaic described in \S\ref{sec:PAH} by correcting each pixel of the feature maps based on the spectral response function and the feature shape. To implement the correction, a continuum image was first calculated for the feature from a linear interpolation between nearby continuum wavelengths. This image was subtracted from the individual L2 exposures to construct a line intensity image. Then for each pixel in the image, the pixels were corrected using a factor based on their central wavelength, the pixel response function, and a model of the PAH emission feature shape based on an average spectral shape common to many Galactic sources \citep[Type 1;][]{1991Tokunaga}. This adjusted the feature strength to what it would be if it had been on the center line wavelength, assuming that the observed feature has the same spectral shape as the model. Once this correction has been applied to the L2 exposures, they were mosaicked to produce a corrected PAH line emission map. An illustration of the effect of the correction is shown in Figure\,\ref{fig:BandCorr}.

\begin{table}[]
    \centering
    \caption{Correction Factor Values}
    \begin{tabular}{|c|cc|}
    \hline
    \hline
    Feature & Center & Max\\
    \hline
     PAH    &  1.95 & 4.24\\
     \hto   &  1.01 & 1.13\\
     \co2 &  1.44 & 2.99\\
     \hline 
    \end{tabular}
    
    \label{tab:corrfac}
\end{table}

The correction values at the center of the line and the maximum correction values are listed in Table\,\ref{tab:corrfac}. The \hto\ line is wide compared to the SPHEREx resolution at 3.05\,\micron, so the variation of the correction is small in the wavelength range included in the spectral line mosaic. The PAH line has a fairly sharp peak, therefore the factor needed to correct the observed to actual line flux is $\sim$\,2 in the line center and is larger near the edges of the band. The CO line is narrower than the other ice absorption lines and in CygX is not as deep as the \hto\ and \co2, and also not sampled well enough in this first set of observations for us to construct similar high-quality corrected optical depth maps. 

\subsection{SPHEREx Spectra}
\subsubsection{L3 Spectra}
The L3 pipeline operated by the SSDC produces spectrophotometry of objects in the SPHEREx reference catalog \citep{2026Yang}. The photometry is based on {\it The Tractor} image modeling code \citep[][]{2016aLang, 2016bLang} which carries out forced photometry at each catalog position using a model of SPHEREx's point source response function (PSF) developed from flight data to extract point sources from the L2 images. The SSDC has also provided a photometry tool\footnote{\url{https://irsa.ipac.caltech.edu/applications/spherex/tool-spectrophotometry}} to enable a user to extract spectra at specific positions from the L2 images, which can be used by the public until the High Reliability Source Catalog is released by the SSDC (see the data products release schedule\footnote{\url{https://spherex.caltech.edu/page/data-products}}
). 
\label{sec:spectra}

The completeness of any spectrum presented here depends on how many SPHEREx L2 images have been obtained of a given sky position at this point in the SPHEREx mission. For most fully-covered areas in this initial dataset, $\sim$100 spectral samples are available from the L2 images. Based on simulations of the survey strategy for the targets in the SPLICES list, we expect most objects in the Galactic plane will eventually be spectrally sampled $\sim$600 times over the course of the mission.

As mentioned above, SPHEREx's spectral resolving power ranges from $R \sim 35$--130 which is too coarse to resolve narrow lines such as those from H or \h2. The spectra  underestimate the peak flux from these unresolved lines, but the integrated flux from these lines should be accurately measured when the pixel response function is taken into account. The appearance of the spectra here will also be affected by the wavelengths sampled in this limited dataset. Some sources will have been observed with the source position on the array nearly centered on the peak wavelength for a particular spectral feature, but some sources may not be well-centered depending on where the source happened to fall on the array. The spectra that happen to sample near the peak of the feature will appear to have stronger line emission compared to a source where the sampling was off the peak of the line emission. This effect will be somewhat mitigated later in the mission as more spectral samples of each source are accumulated. We are also developing a forward-modeling approach that uses the pixel response function to recover intrinsic line profiles to properly infer the peak flux and integrated flux of emission and absorption lines, which will be utilized in future analyses.

In the lowest resolution range from 2.4 - 3.8\,\micron\ in band 4,  narrow lines such as from Br\,$\beta$ are diluted more severely and difficult to detect. However, the broad \hto\ ice absorption line centered near 3\,\micron\ and the 3.28\,\micron\ PAH line are marginally resolved and the peak intensities are closer to the actual maxima that would be observed with sufficient resolution. The \co2\ and CO ice absorption features are located in bands 5 and 6 where the \spherex\ spectral resolution is $R\sim110-130$ to better sample these features.

The limited spectral resolution will affect the determination of the peak optical depth and the integrated optical depth of the ice absorption features. Since the optical depth is calculated as the natural logarithm of the ratio of the continuum to line intensity, the peak optical depth will be underestimated for unresolved lines and the measured values cannot be simply summed to determine the integrated optical depth and column densities of the ice species. For the Ices investigation pipeline, we have calibrated this effect using simulated observations of spectra with known optical depths (see V. Tolls et al. 2026, in preparation) to determine the correction factors necessary to convert the measured values to the true optical depth in the SPHEREx spectra. For the spectra presented below, we have not applied any correction factors, so the optical depths shown are lower than the true values, and this especially affects the narrow \co2 and CO absorption features. However, spectra of different point sources can be compared to detect the variation in the relative feature strength, as shown in \S\ref{sec:icespec}.

\subsubsection{Aperture Photometry}
The L3 photometry and online tool are not appropriate for extracting extended emission from arbitrary regions, so we developed an aperture photometry tool to produce spectra of extended emission, and also sources detected by SPHEREx but not in the reference catalog or objects in regions with complicated backgrounds and crowded fields. This tool uses the reprojected image slices produced in the mosaicing process. To produce a spectrum, the user chooses the aperture location and size, and also optionally selects a region off-source to use as a background reference. The source density in Galactic‑plane mosaics is high, and the fields often contain complex extended emission. For this reason, we chose not to use a standard annulus around each source. Instead, the user selects a nearby region representative of the local background to obtain an appropriate background estimate. A median is taken of the background region which is used to subtract from the on-source aperture to photometer the source at each wavelength from the individual slice images and assemble a spectrum. If the source aperture contains a pixel marked as bad in an image slice, that photometric point is not included in the spectrum. The center wavelength of the pixel response is also read from the image slices, so the program records the wavelength of each measurement in order to accurately place it in the resulting spectrum.


\subsection{Herschel SPIRE Mosaics}
\subsubsection{Mosaic Construction}
Herschel SPIRE (Spectral and Photometric Imaging Receiver; \citealt{2010Griffin}) imaging data were used to construct column density and dust temperature maps to compare to the SPHEREx mosaics. This step required building large mosaics centered on the CygX region at 250, 350 and 500 $\mu$m, starting from the highest level data products  available in the Herschel Science Archive (HSA{\footnote{https://archives.esac.esa.int/hsa/whsa/}}) at ESA. 

The final mosaics, spanning 11 deg $\times$ 5 deg were generated using the publicly available 
Montage\footnote{http://montage.ipac.caltech.edu/} software. 
SPIRE Level-3 data (processing version 14.2.1) from an initial set of 53 obsids (mostly from programs {\em{KPGT$\_$fmotte$\_$1}}, {\em{OT2$\_$smolinar$\_$7}}, {\em{GT2$\_$pandre$\_$5}} and {\em{OT2$\_$jhora$\_$2}}) were used. The original Level 3 data products are in azimuthal (ARC) coordinate projection, which is not compatible with Montage. Therefore, prior to mosaicing, the Level 3 data were re-projected on a tangent (TAN) coordinate grid, using the Python 
{\it{reproject{\footnote{https://reproject.readthedocs.io/en/stable/}}}} package. Since the archival SPIRE Level-3 data products are absolutely calibrated through Planck \citep{2011PlanckCollab} ancillary data, no background correction needed to be applied.  The final SPIRE mosaics have a pixel size of 6\arcsec (250 $\mu$m), 10\arcsec (350 $\mu$m) and 14\arcsec (500 $\mu$m). 

\subsubsection{Column Density and Temperature Maps}
\label{sec:herschel_map_method}
The $N_\mathrm{H_2}$ ($\mathrm{cm}^{-2}$) and $T_\mathrm{d}$ (K) maps were derived from SPIRE data at 250, 350, and 500\,$\mu$m with beam full width at half maximum (FWHM) of 18.1\arcsec, 24.9\arcsec, and 36.4\arcsec, respectively. All maps were smoothed to the angular resolution of the 500\,$\mu$m band and reprojected onto the same grid as the SPHEREx mosaic images. 

We performed pixel-by-pixel spectral energy distribution fitting assuming optically thin, single-temperature dust emission described by a gray-body model, following the methodology adopted in the Herschel Gould Belt Survey \citep[e.g.,][]{2010Konyves, 2014Roy}. We adopted a power-law dust opacity per unit (dust+gas) mass, $\kappa_\nu = 0.1 (\nu / 1000\,\mathrm{GHz})^{\beta}\ \mathrm{cm}^2\,\mathrm{g}^{-1}$, with dust emissivity index $\beta = 2$ \citep{1983Hildebrand}. The column density $N_\mathrm{H_2}$ is computed assuming $\mu = 2.8$ per hydrogen molecule, with all hydrogen in molecular form.

\section{Spectral Feature Maps}

The spatial distribution of the optical depth of the strongest ice features of \hto\ (3.05\,\micron) and \co2\ (4.27\,\micron) is clearly revealed in the  SPHEREx maps. The peak optical depth $\tau$ was calculated from mosaics centered on the absorption peak wavelengths. The continua at the line centers $F_{\lambda_C}$ were estimated by linearly interpolating between wavelengths adjacent to the lines, and then calculating $\tau = ln(F_{\lambda_C}/F_{\lambda_I})$, where $F_{\lambda_I}$ is the flux in the image at the ice absorption line center. We present the CygX ice maps for \hto\ and \co2\ in \S\ref{sec:iceabs}.

\begin{figure*}
    \centering
    \includegraphics[height=8.5in]{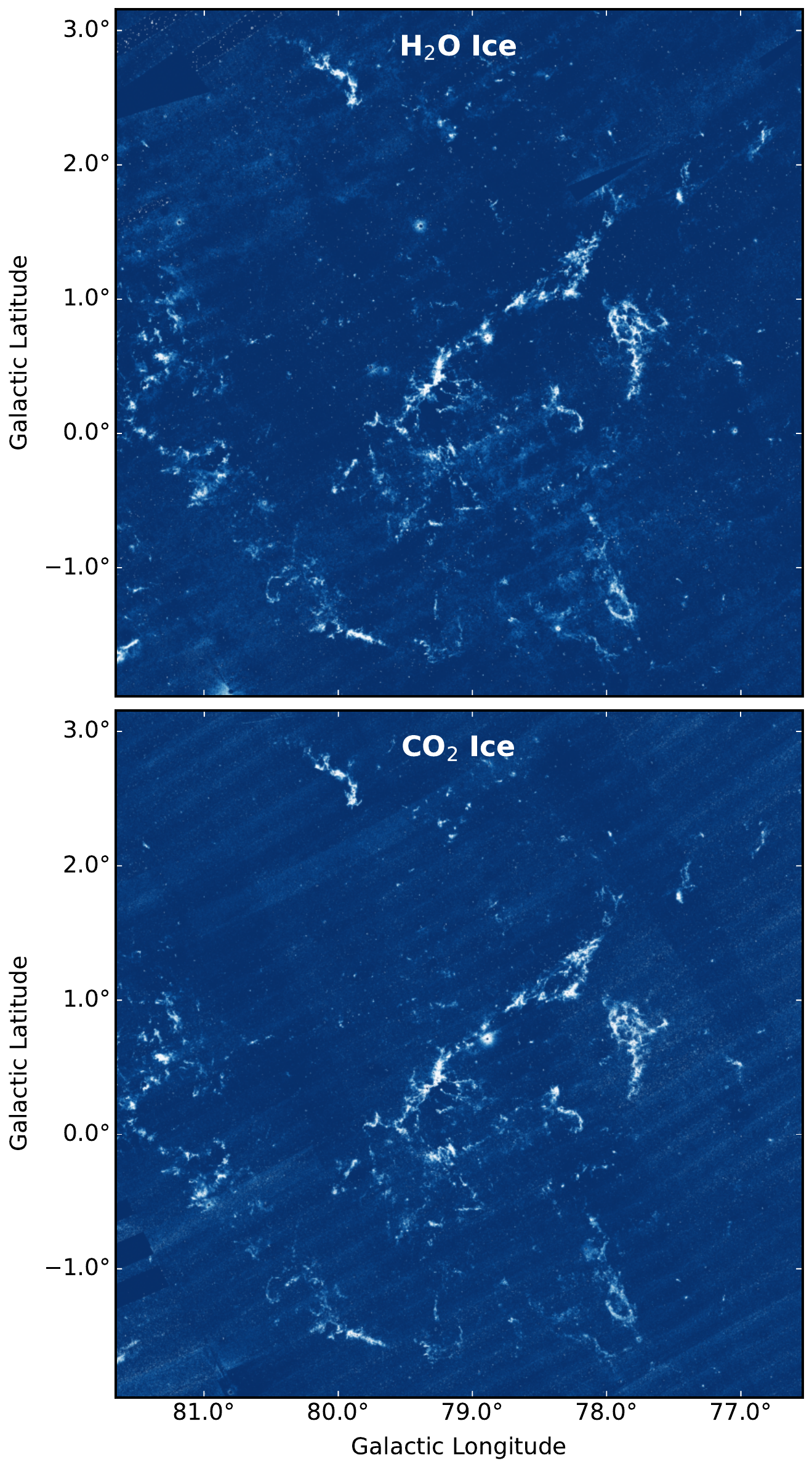}
    \caption{Ice peak optical depth maps in \hto\ ice (top) and \co2\ ice (bottom) for the same region in CygX shown in Figure\,\ref{fig:colorcyg}. The optical depth is shown using a color scale with white corresponding to $\tau=0.75.$ The distributions of the two ice species is similar, but there are some differences in relative absorption depths as shown in Figure\,\ref{fig:icezoom}. Some residual ``banding'' or striping is visible in these images, see the discussion in \S\ref{sec:banding}.}
    \label{fig:iceabs}
\end{figure*}
\begin{figure*}
    \begin{flushright}
    \includegraphics[width=0.985\linewidth]{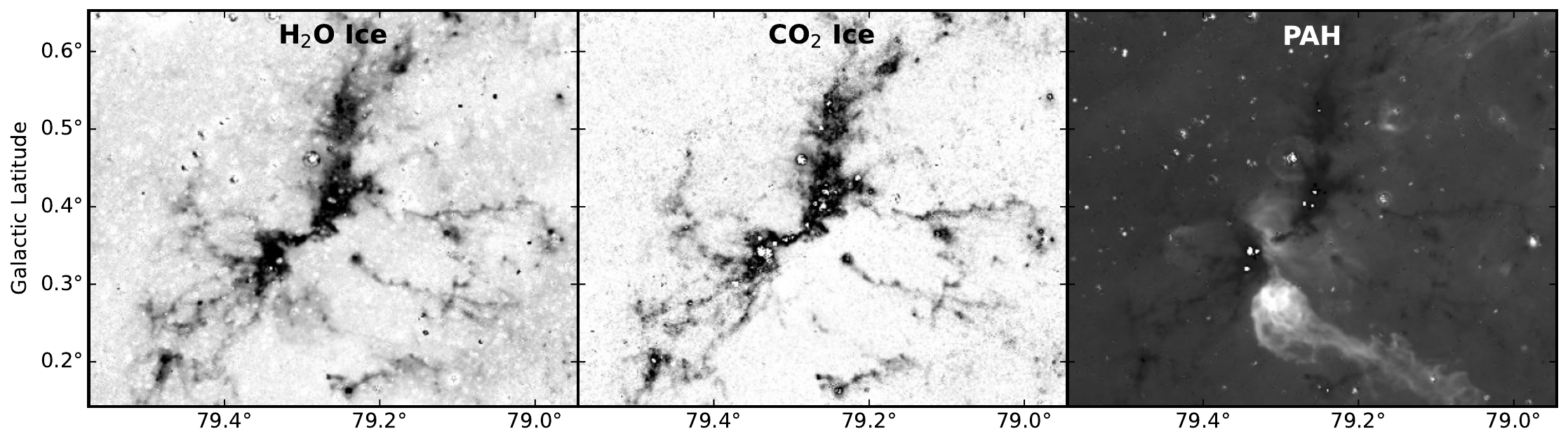}
    \includegraphics[width=0.985\linewidth]{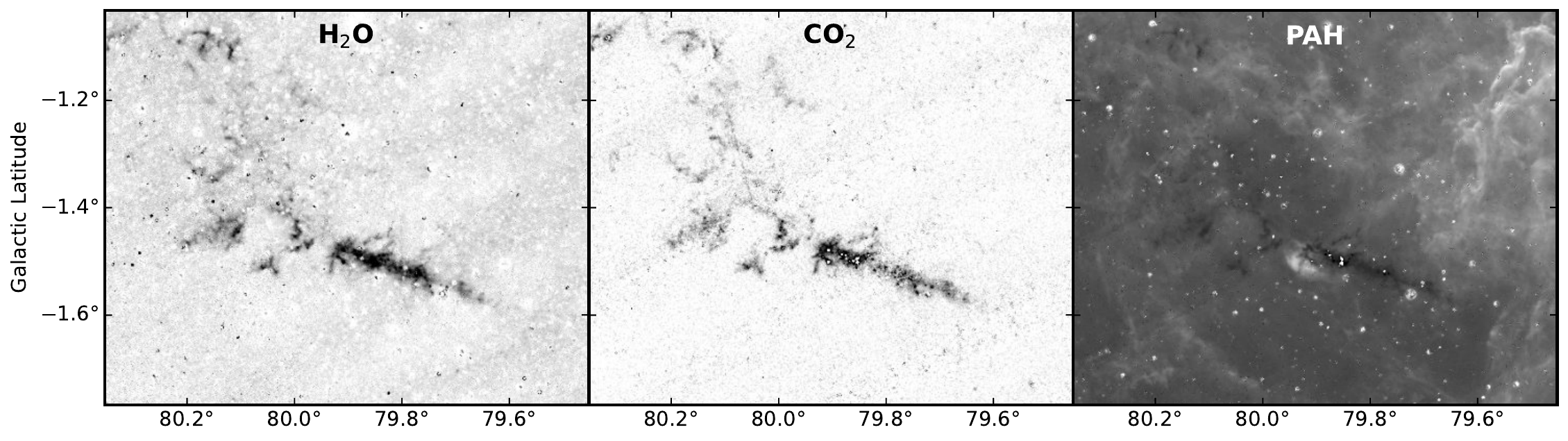}
    \includegraphics[width=0.985\linewidth]{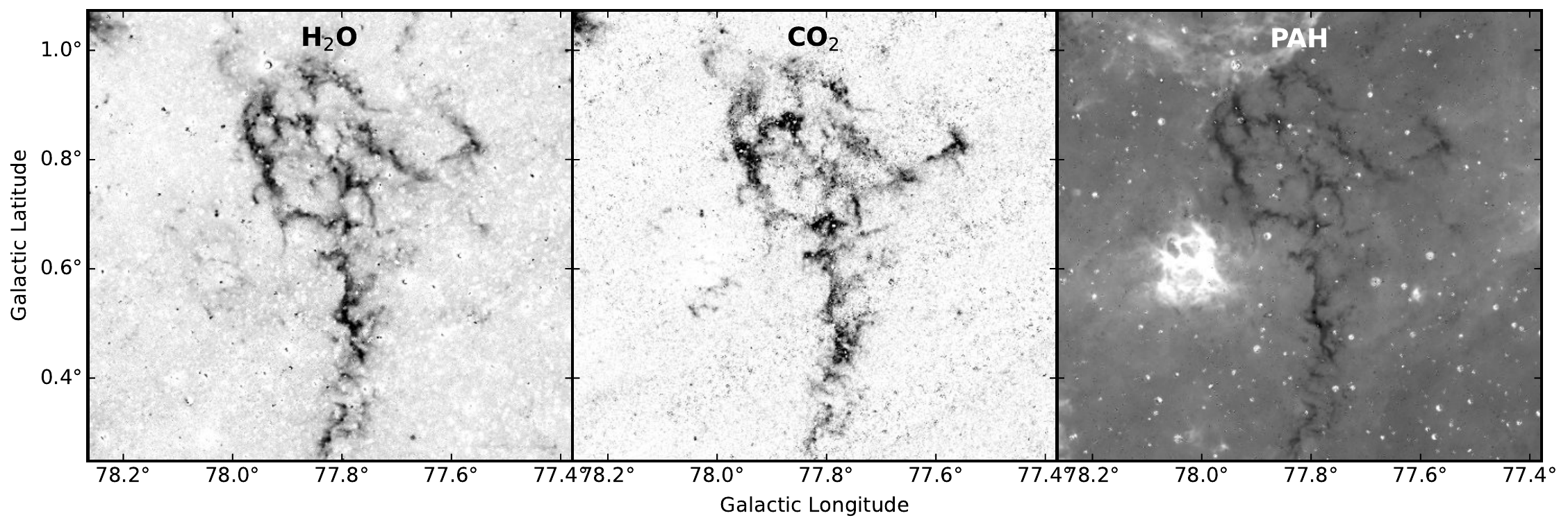}    

    \caption{Three selected regions of \hto\ ice and  \co2\ ice absorption line images from Figure\,\ref{fig:iceabs}, along with the corresponding section of the PAH line image. The ice absorption images show the peak optical depth in the feature in inverse grayscale, ranging from 0 to 1.1. The PAH images show lower flux regions as black and higher flux as white. The top images zoom in on a region near the center of the CygX mosaic shown in Figure\,\ref{fig:colorcyg}, and includes DR\,15 which is the bright \ion{H}{2} region near the bottom of the PAH image. The middle images include the dark cloud LDN\,896. The bottom row shows the dark cloud complex that is slightly to the right of center in Figure\,\ref{fig:colorcyg}, cataloged as the 2MASS dark clouds Dobashi 2362 and 2367 \citep{2011Dobashi}, and known from Spitzer/IRAC imaging as being composed of a network of filamentary dark clouds \citep{2020Pari}}.
    \label{fig:icezoom}
        \end{flushright}
\end{figure*}

In a similar way,  emission line maps can be constructed by estimating the continuum level by linearly interpolating from adjacent continuum spectral points and subtracting the continuum from the mosaic centered on each emission line wavelength. Examples are shown in \S\ref{sec:PAH} for the 3.28\,\micron\ PAH feature, \S\ref{sec:HII} for hydrogen recombination lines, and \S\ref{sec:H2} for \h2\ emission.

\begin{figure*}
    \centering
    \includegraphics[width=0.95\linewidth]{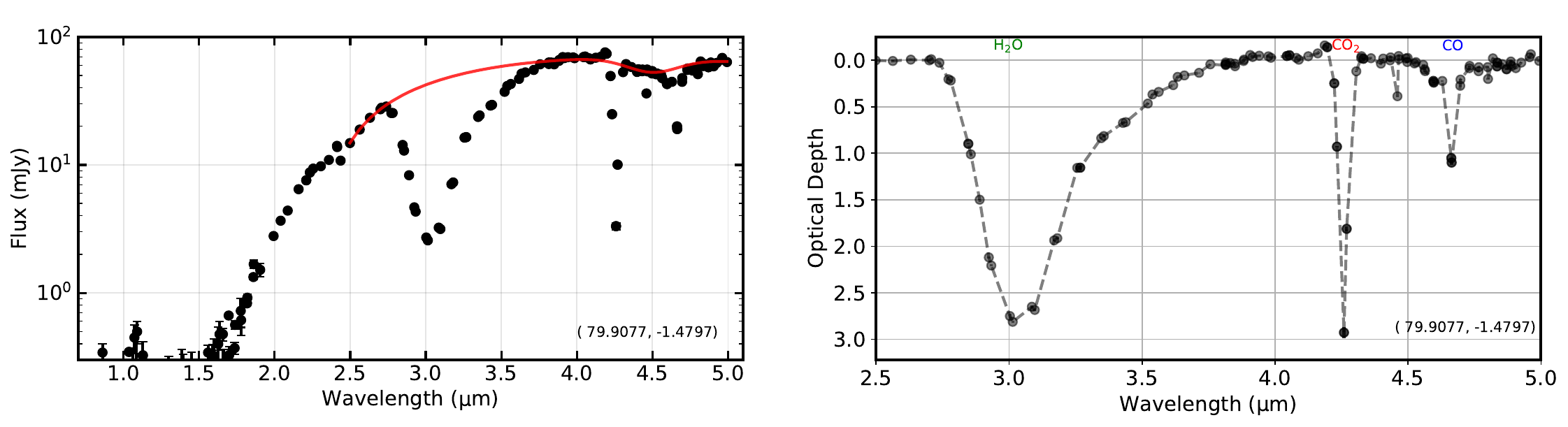}
    \includegraphics[width=0.95\linewidth]{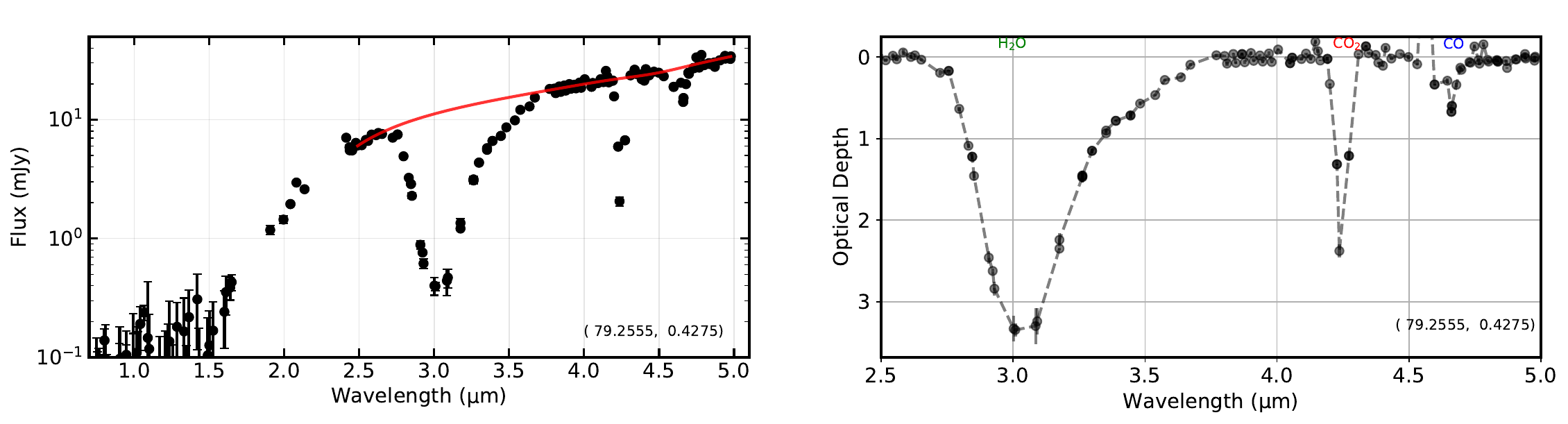}
    \includegraphics[width=0.95\linewidth]{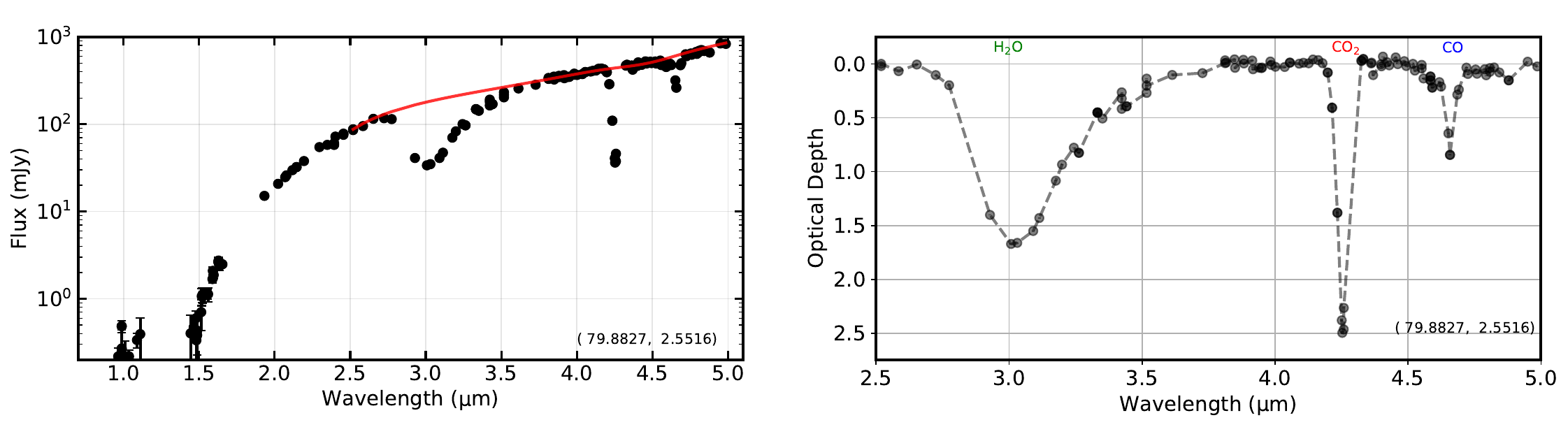}
    \includegraphics[width=0.95\linewidth]{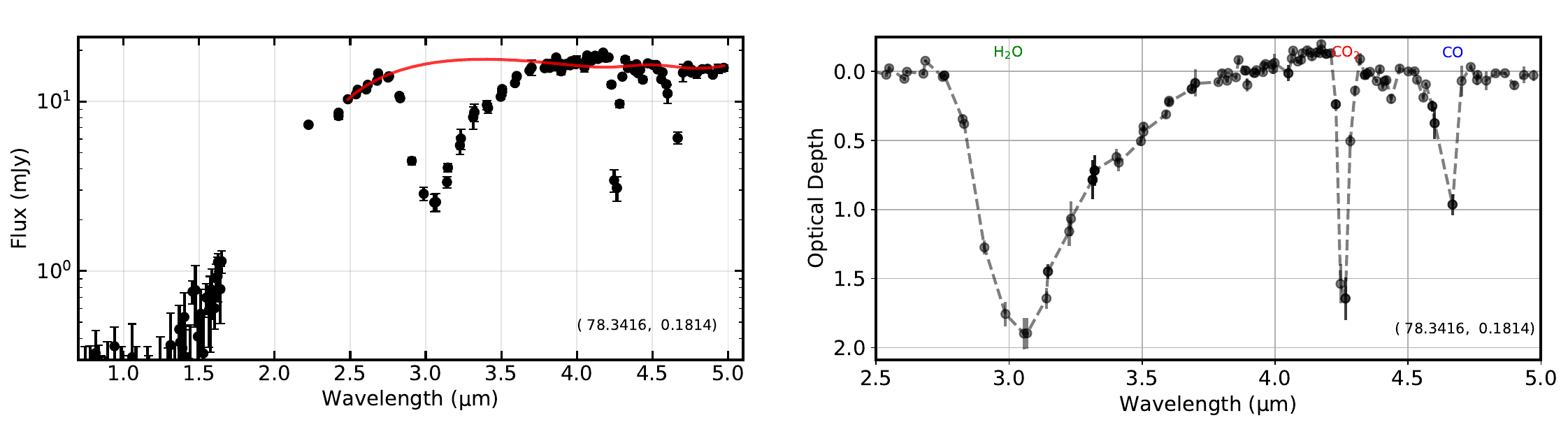}
    \caption{Spectra and optical depths of sample sources in the CygX dark clouds obtained with aperture photometry using the L2 images. The Galactic coordinates ($l, b$) are shown in the lower right corner of each panel. The plots on the left show the photometry along with the continuum fit for wavelengths $>$2.5\,\micron\ plotted in red. The plots on the right show the calculated optical depths for the spectra on the left after modeling and removing the broad \hto\ ice feature at 4.6\,\micron. The dashed line connects the data points and is not a fit to the absorption depth. Note that the relative strengths of the features vary among the different lines of sight throughout the clouds. }
    \label{fig:apphot}
\end{figure*}

\subsection{Ice Absorption Maps}
\label{sec:iceabs}
The absorption maps we constructed using the technique described above show the distribution of the ice absorption feature over the region being displayed. The diffuse emission from the ISM and unresolved faint sources is absorbed by the clouds and they appear dark against the background of the Galactic plane. These feature maps, however, do not measure the true peak absorption depth when the peak $\tau \sim 1$ or more. 

The reason is that since the flux level of the background emission passing through the cloud at the line center is lower than the SPHEREx detection limit, the measured flux is set by the amount of foreground emission, while outside of the ice feature the continuum  is set by the sum of the foreground emission and the background emission incident on and passing through the cloud. Thus the measurement is contaminated by the foreground emission at the line center and the optical depth cannot be accurately determined, regardless of the actual amount of ice along the line of sight. This constraint does not apply to lines of sight towards bright stars, where the starlight from sources behind the cloud can be detected even at the line center. In those cases the foreground emission is relatively much fainter and is subtracted when performing photometry, assuming its contribution to the flux on the point source can be estimated from nearby pixels in the mosaic.

The \hto\ and \co2\ ice optical depth images of the CygX region are shown in Figure\,\ref{fig:iceabs} for the same area covered by Figure\,\ref{fig:colorcyg}.
Some of the ice absorption structures extend over several degrees. The distribution of the ice absorption roughly follows the extent of the dark clouds seen in Figure\,\ref{fig:colorcyg}, but the features are much easier to see in the ice feature images because the emission from stars and the ISM has been suppressed in the optical depth calculation. The spatial distribution of the \hto\ and \co2\ optical depth appear very similar in these wide-field images. In Figure\,\ref{fig:icezoom} we enlarge the images of three selected regions to compare their spatial structure in detail. In many places along the cloud there are dark spots from background stars that are shining through the clouds and exhibit much higher optical depths since they are sampling the column density of ice along the full line of sight through the cloud at that position. We present the spectra of some of these sources in the following section.

\subsection{Ice Absorption Spectra}
\label{sec:icespec}

The spectra of stars that lie behind absorbing clouds can be used to accurately determine the peak $\tau$ along lines of sight through the cloud. By construction, these objects are sufficiently bright to be detected, although we must check that the features do not suffer from saturation at their peaks.  The adjacent backgrounds can be estimated and subtracted to eliminate the effects of foreground emission. Many individual lines of sight are needed to gain a full picture of the cloud, which will be possible only after fully-sampled spectra are obtained at the completion of the primary mission.  At the present early mission stage, the spectra are sparsely sampled.  Many of these background stars are nonetheless visible as dark spots along the filaments in Figure\,\ref{fig:icezoom} because of their higher optical depths.  With our aperture photometry tool we extracted representative spectra from the L2 images. Examples are shown in Figure\,\ref{fig:apphot}.

The spectra confirm that in addition to extinction due to dust, the dark clouds have significant amounts of \hto, \co2, and CO ice absorption. The relative feature strengths  and their peak optical depths both vary along different lines of sight through the clouds. The peak of the broad \hto\ feature is fairly accurately measured in these spectra, but the \co2\ and CO feature depths are not fully sampled due to their narrow widths relative to the SPHEREx resolution (\S\ref{sec:spectra}). When deriving integrated optical depths from the spectra, a correction factor must be applied to these features to convert from the measured to the actual values. We are in the process of deriving the correction factors based on data from higher-resolution JWST measurements \citep[e.g.,][]{2023McClure,2024Bergner} using the SPHEREx ices simulator \citep{Crill_2025, 2026Tolls} which will be described in a future paper.
\begin{figure*}
    \centering
    \includegraphics[width=0.519\linewidth]{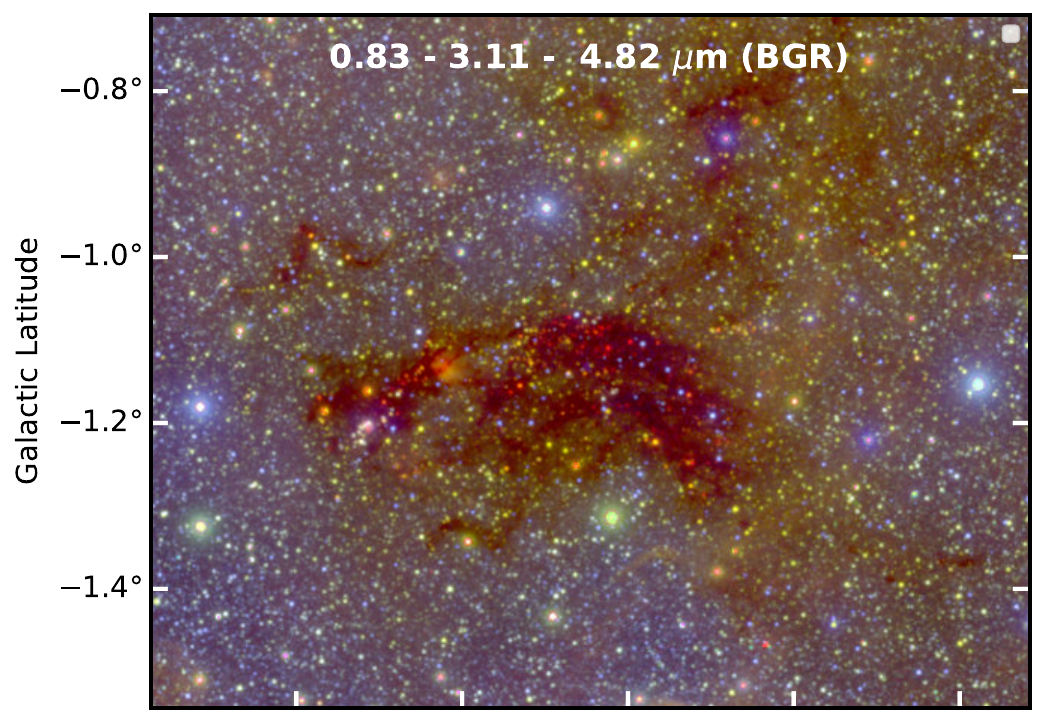}
    \includegraphics[width=0.452\linewidth]{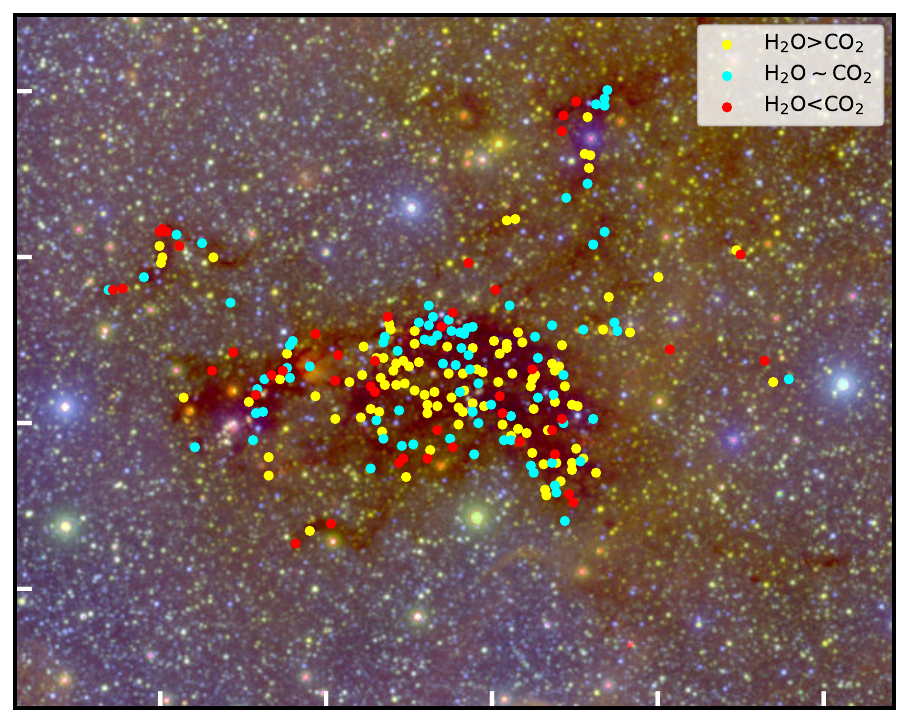}
    \includegraphics[width=0.518\linewidth]{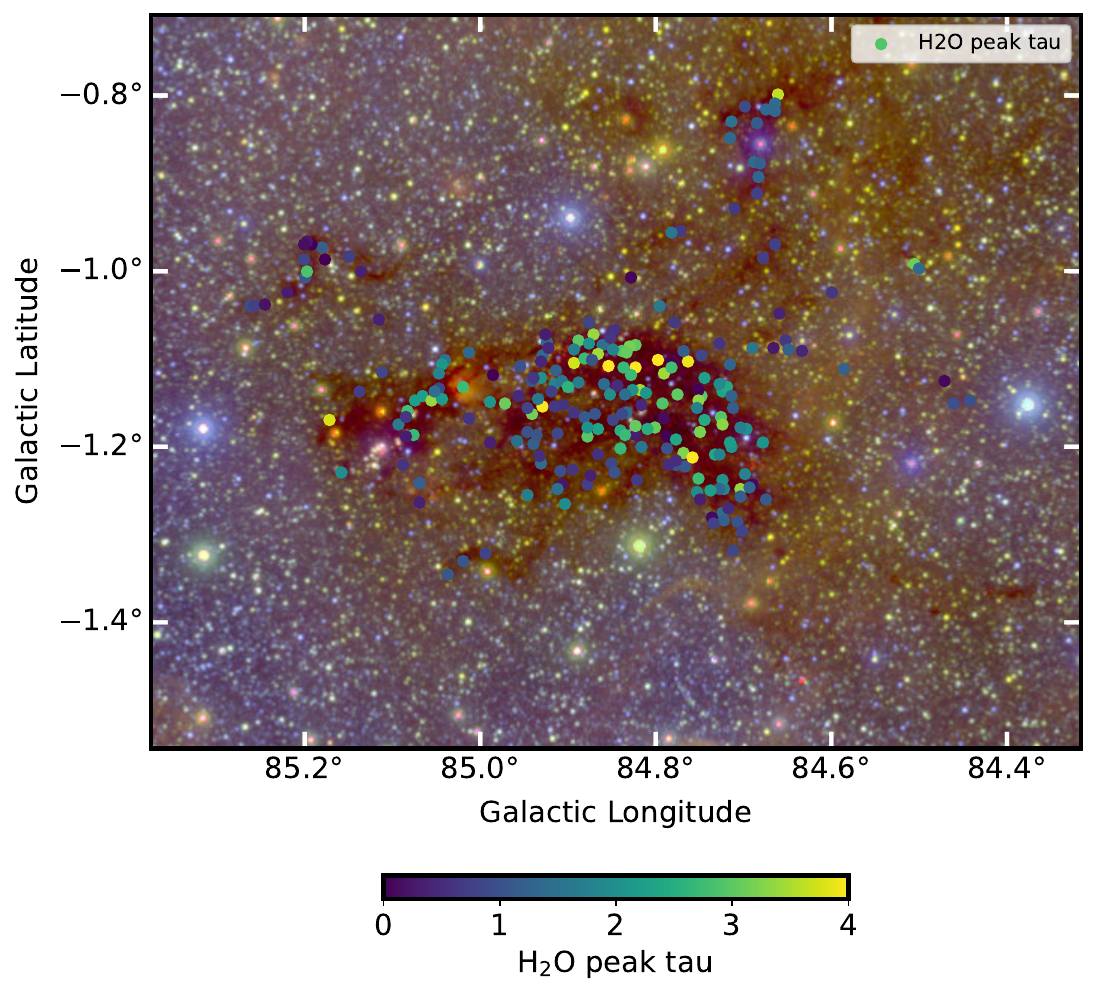}
    \includegraphics[width=0.453\linewidth]{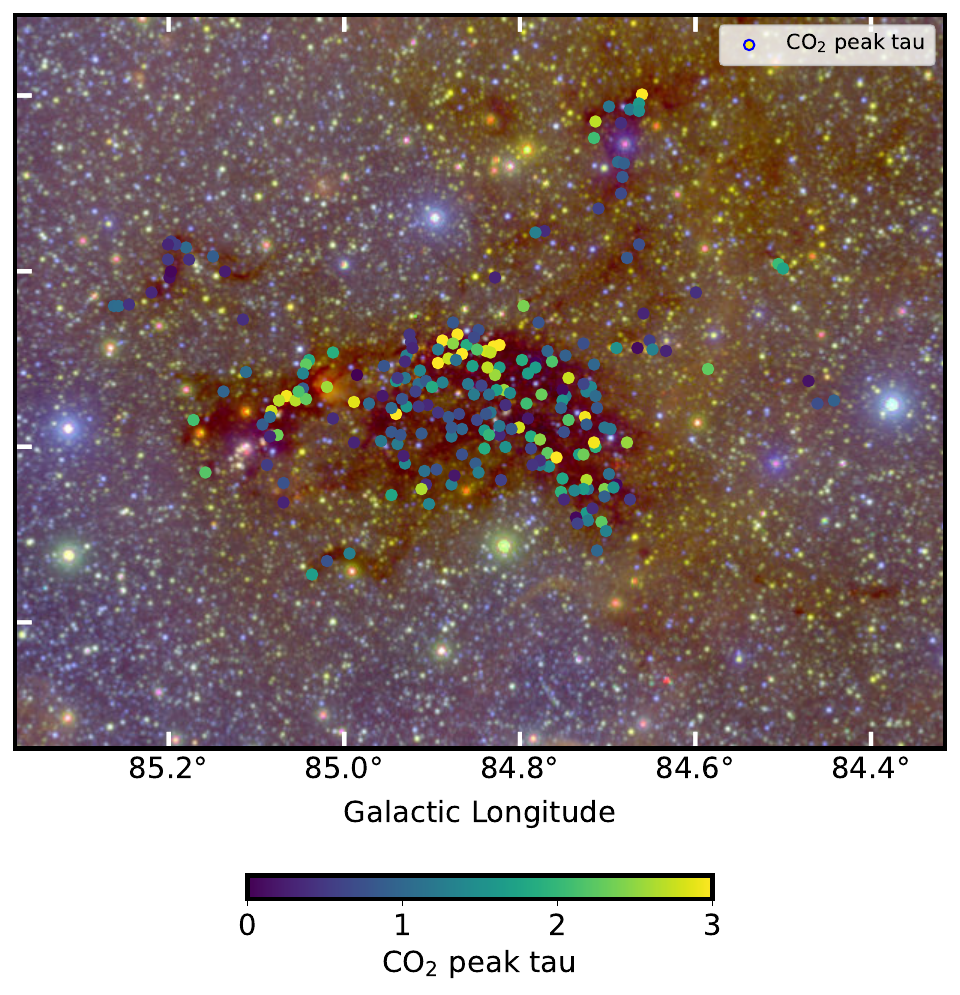}
    \caption{The SPHEREx mosaic of the dark cloud LDN\,935. All panels show the same 3-color image as the upper left panel, using wavelength ranges centered at 0.83, 3.11, and 4.82\,\micron\ for blue, green, and red, respectively. The other three panels show the positions of the point source spectra extracted, with color-coded circles. In the lower left panel, the points are coded according to the peak $\tau$ of the 3.05\,\micron\ \hto\ ice absorption feature, with the scale shown below the plot. In the lower right panel, the points are color-coded according to the peak $\tau$ of the 4.27\,\micron\ \co2\ ice feature. The upper right figure, the points are coded according to whether the peak $\tau$ of the \hto\ feature is 20\% or more larger than the \co2\ peak $\tau$ (yellow), the two features are within 20\% of each other (cyan), or the \co2\ feature is more than 20\% larger than that of the \hto\ feature (red).}
    \label{fig:NAmN}
\end{figure*}
\subsection{LDN\,935 Images and Spectra}
A three-color image of LDN\,935 is shown in each of the four panels of 
Figure\,\ref{fig:NAmN}.
We extracted spectra for a set of 231 sources in that dark cloud using aperture photometry, and their positions are indicated in three of the panels. We performed continuum fitting and calculated the optical depth spectra, similar to those displayed in Figure\,\ref{fig:apphot}. Because of the large number of spectra, we do not plot them all separately here, however instead we extracted the peak optical depth $\tau_{peak}$ in the 3.05\,\micron\ \hto\ and 4.27\,\micron\ \co2 ice absorption features and summarize the results in the figure. The points indicate the relative strength of the absorption, as indicated in the caption.

One region of high optical depth in both \hto\ and \co2\ ice is near the center of the image, where we see high $\tau_{peak}$ in both features denoted by the yellow points in the lower panels of  Figure\ \ref{fig:NAmN}. The upper right panel shows that at this location there are a cluster of cyan points indicating that the $\tau_{peak}$ values are similar. According to the analysis of CO molecular cloud structure and dust mapping distances using Gaia data by \citet{2021Kong}, this high $\tau$ region may be associated with a part of the cloud they call ``F-7'' which is likely behind  W80  and the other parts of LDN\,935 that lie in front of the bubble. The upper right panel shows a large fraction of the region with \hto\ $\tau_{peak}$ $>$ \co2, as indicated by the yellow circles in the central part of the image. There are also several places where \co2\ is greater than the \hto\ peak, such as the set of points in the upper right of the image near ($l=85\fdg2, b=-1\fdg0$). These sources sample lines of sight through a feature that \citet{2021Kong} call the ``Dark worm'' which they determine is located in front of the W80 bubble. This may be an especially dense filament, or it may be shielded from the UV radiation from the O3.5 star by the other parts of the nebula.

The SPHEREx imaging of CygX and LDN\,935 presented here demonstrate the potential of the mission for wide-field spectral imaging, and in particular, the ability to combine point-source spectroscopy with diffuse spectral imagery.  A more complete analysis of these fields including determining the relative positions of the molecular clouds and the background sources will be presented in a subsequent paper. 
\begin{figure*}
    \centering
    \includegraphics[width=0.99\linewidth]{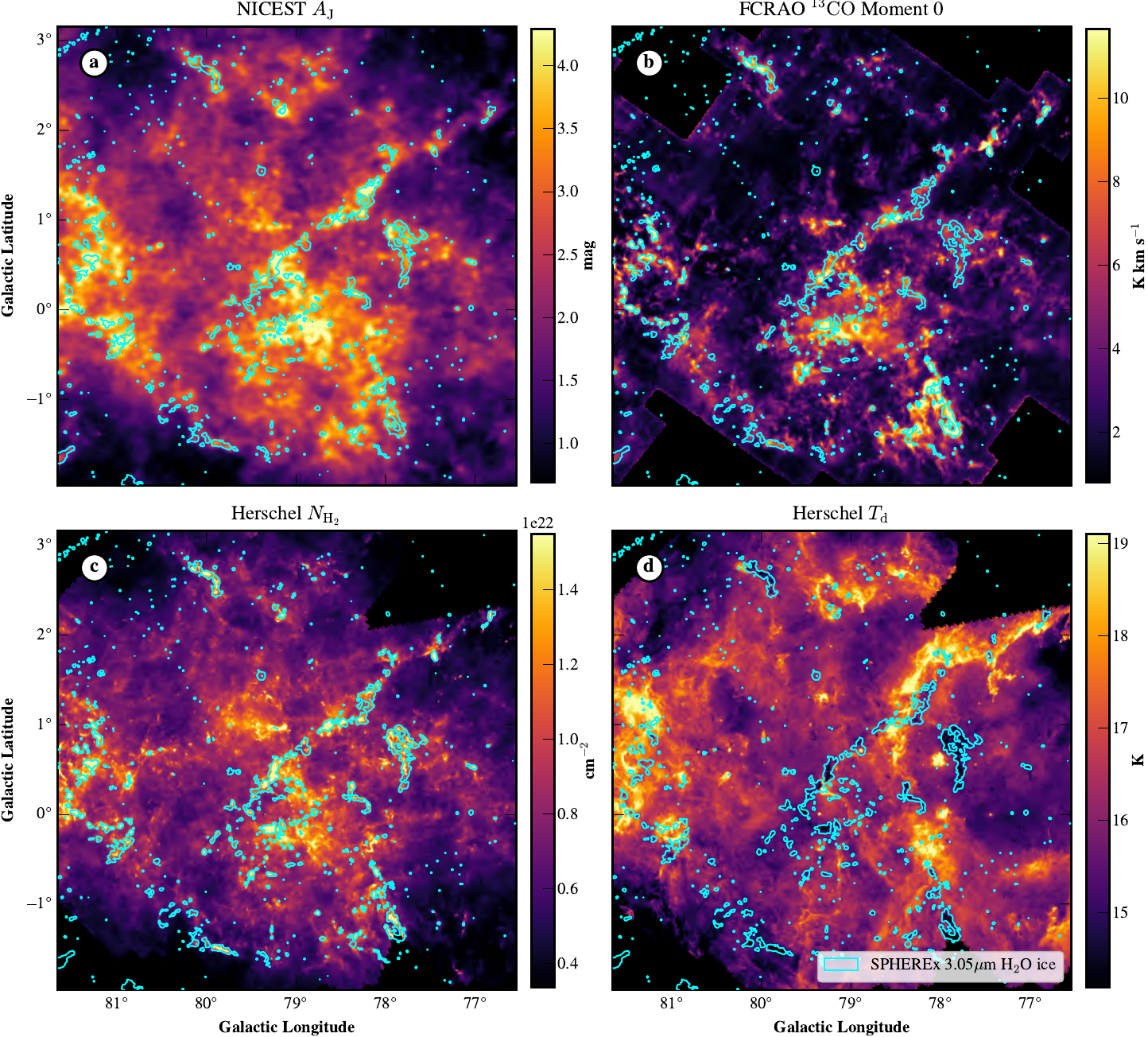}
    \caption{ Comparison of various tracers of dust, gas, and extinction toward the CygX region. Cyan contours in all images indicate $\mathrm{H_2O}$ ice peak optical depths of 0.25, derived from the $\mathrm{H_2O}$ ice map smoothed with a Gaussian kernel of $\sigma \sim$ 31\arcsec\,(5 pixels). (a) 2MASS NICEST $A_{\mathrm{J}}$ extinction, with a resolution of 3\arcmin. (b) FCRAO $\mathrm{^{13}CO}$ (1--0) Moment~0 map (beamsize 45\arcsec\ FWHM) integrated over $-10$ to $+20\,\mathrm{km\,s^{-1}}$, corresponding to the ensemble of star-forming molecular clouds directly associated with the Cyg~OB2 cluster \citep{Schneider2006}. (c) Herschel-derived $N_{\mathrm{H}_2}$ column density. (d) Herschel-derived dust temperature, $T_{\mathrm{d}}$. Both Herschel images are at a resolution of 36.4\arcsec.
}
    \label{fig:Dustmaps}
\end{figure*}

\begin{figure*}
    \centering
    \includegraphics[width=0.99\linewidth]{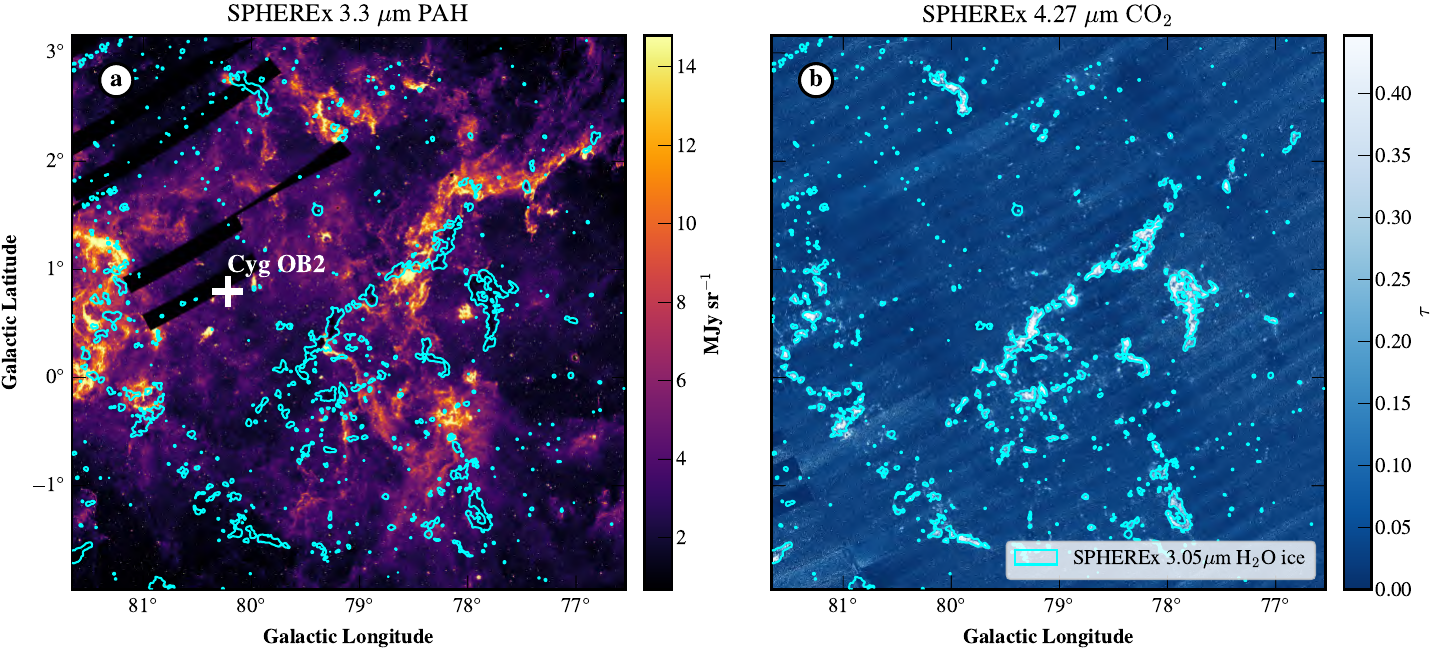}
    \caption{ Comparison of the PAH emission and ice species absorption toward the CygX region. The SPHEREx maps are constructed at a pixel scale of 6\parcs{15}. Cyan contours in all images indicate $\mathrm{H_2O}$ ice peak optical depths of 0.25, derived from the $\mathrm{H_2O}$ ice map smoothed with a Gaussian kernel of $\sigma \sim$ 31\arcsec\,(5 pixels). a) SPHEREx 3.28\,$\mu$m PAH emission. The white cross marks the position of Cyg~OB2. (b) SPHEREx 4.27\,$\mu$m $\mathrm{CO_2}$ ice peak optical depth. 
}
    \label{fig:PAHandIce}
\end{figure*}

\subsection{Comparison to Dust Extinction and Emission maps} \label{subsec:dustmaps}

We compare several dust and ice tracers in Figures\,\ref{fig:Dustmaps} and \ref{fig:PAHandIce}. The NICEST extinction map is adopted from \cite{Juvela2016} (with a resolution of 3\arcmin), 
while the $^{13}$CO (1--0) map is taken from \cite{Schneider2007, 2010Schneider}, obtained with the FCRAO 14\,m telescope and a beam FWHM of 45\arcsec. The $\mathrm{H_2}$ column density ($N_\mathrm{H_2}$) and dust temperature ($T_\mathrm{d}$) maps were derived from Herschel SPIRE observations at 250, 350, and 500\,$\mu$m. All maps are smoothed to a common angular resolution of 36.4\arcsec\ and reprojected onto the SPHEREx mosaic grid. Details of the derivation of the Herschel-based $N_\mathrm{H_2}$ and $T_\mathrm{d}$ maps are presented in \S\ref{sec:herschel_map_method}.

Both \hto\ (cyan contours) and \co2\ ice (panel~\ref{fig:PAHandIce}b) optical depth maps broadly follow the dust tracers shown in Figure\,\ref{fig:Dustmaps}. For \hto\ in particular, the Spearman’s rank correlation coefficients are 0.461, 0.388, and 0.272 with the FCRAO $^{13}$CO, NICEST extinction, and Herschel column density maps, respectively; all maps were convolved to the coarsest $3\arcmin$ resolution and masked to the same region for comparison. This trend supports the expectation that ice formation is most efficient in dense, well-shielded regions \citep{Whittet2001, 2015Boogert}, especially the densest clouds traced by $^{13}$CO. 

Regions showing strong ice absorption but weak $^{13}$CO emission may be explained by gas-phase CO depletion onto grain surfaces in cold, dense environments \citep{1999Caselli, jelee2003, jelee2004}. A comparison with the dust temperature map (panel~\ref{fig:Dustmaps}d) further shows that the ice contours preferentially trace lower-temperature lines of sight, indicating that the detected ice resides in cold regions. In contrast, the 3.28 $\mu$m PAH emission (panel~\ref{fig:PAHandIce}a) closely follows the dust temperature distribution (panel~\ref{fig:Dustmaps}d), with enhanced PAH emission in regions of higher dust temperature. This correlation suggests that strong UV radiation both heats the dust grains and excites PAH emission. 


Overall, while the different dust tracers exhibit varying levels of correspondence with the ice maps, all results are consistent with the established picture of efficient ice formation in dense molecular clouds. This agreement holds even over the large spatial extent of CygX, which encompasses multiple giant molecular clouds.

\begin{figure*}
    \centering
    \includegraphics[width=1.0\textwidth]{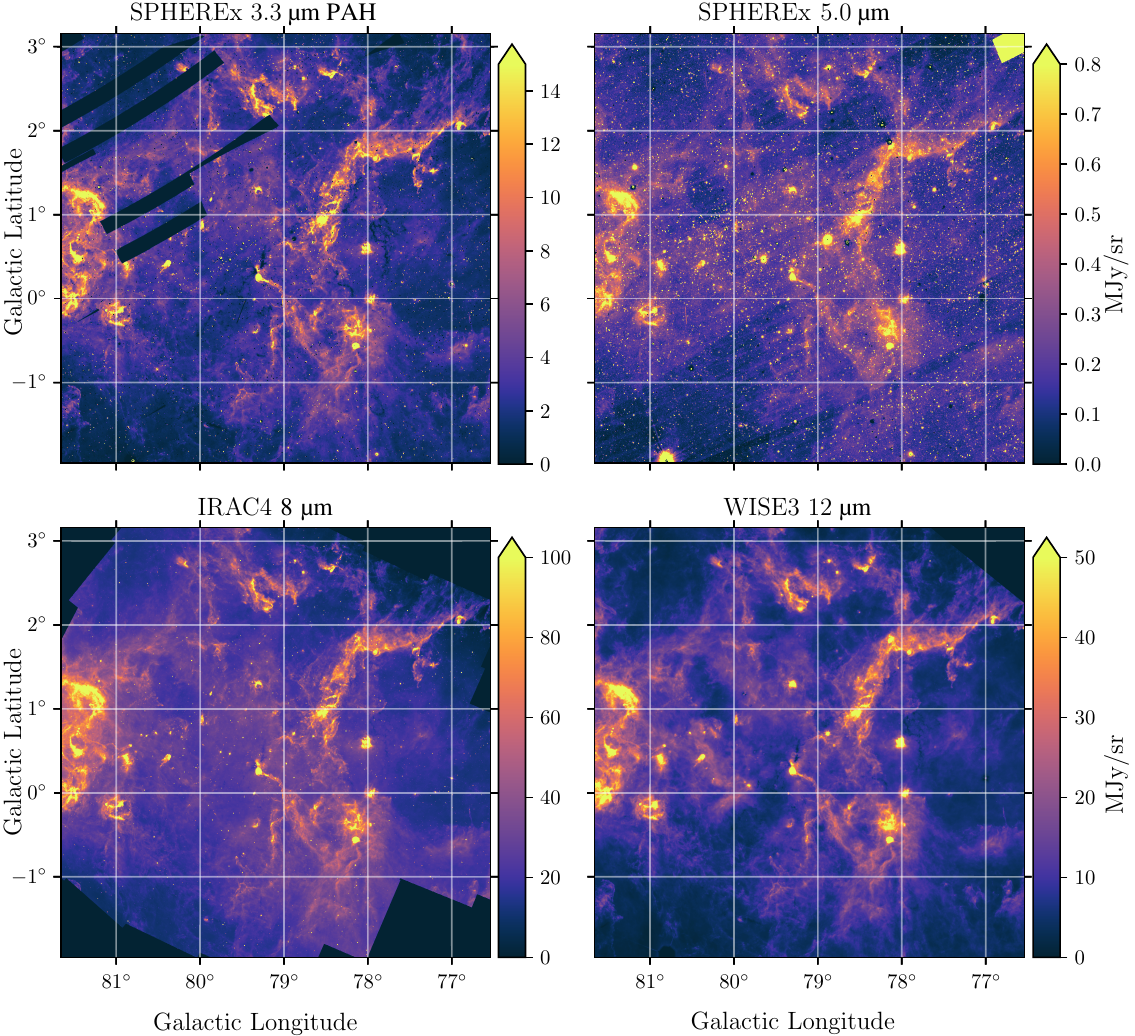}
    \caption{Comparison of the continuum-subtracted SPHEREx 3.28\,\micron\ PAH map (top left), SPHEREx 5\,\micron\ map (top right), Spitzer IRAC Band 4 map \citep[bottom left;][]{DVN/1QMXB8_2021}, and WISE Band 3 map \citep[bottom right;][]{Meisner:2014} in the central part of CygX. The continuum images used for the PAH map were centered on 2.77 and 3.85\,\micron, and a linear interpolation was used to construct a continuum image at 3.28\,\micron. The wavelength range used for the PAH feature is 3.24 - 3.34\,\micron. The banding effect in the PAH map was corrected as described in \S\ref{sec:banding}. The SPHEREx 5\,\micron, IRAC, and WISE maps have had 0.3, 18, and 2~MJy\,sr$^{-1}$ offsets subtracted, respectively. The WISE map has had point sources removed \citep[see][for details]{Meisner:2014}.}
    \label{fig:pah_emission}
\end{figure*}

\subsection{Distribution of 3.28\,\micron\ PAH emission}\label{sec:PAH}

The 3.28 $\mu$m PAH feature is seen prominently in emission throughout the CygX region. To study the relation between PAH emission and other tracers, we have performed a continuum subtraction and correct for banding effects. The resulting PAH emission map is shown in Figures\,\ref{fig:PAHandIce} and \ref{fig:pah_emission}.

Figure\,\ref{fig:PAHandIce} demonstrates that the spatial distributions of PAH emission and ice absorption show poor correlation. This is not unexpected, and several effects may be at play. First, ice exists only in regions that are shielded from UV radiation that can sublimate it. Even if PAHs were present in these highly shielded regions, the lack of UV photons would greatly diminish their emission, particularly at 3.28 $\mu$m  \citep{Draine:2001,  Mori:2012, Rigopoulou:2021, Richie:2025}. Second, any PAH emission located behind the ice on the line of sight would be strongly absorbed by the 3\,$\mu$m \hto\ ice feature. Third, if PAHs indeed undergo rapid growth in molecular clouds \citep{Zhang:2025}, then the 3.28\,$\mu$m emission, which is carried by the smallest PAHs, would be greatly diminished \citep{Lee:2025}. Finally, at such high densities, PAHs may freeze out onto larger grains and thus no longer be able to attain the high temperatures required to emit at 3.28\,$\mu$m.

The lack of spatial correlation between PAH emission and ice absorption is likewise consistent with the longstanding observational studies of the 3.28\,$\mu$m emission feature and tracers of molecular gas \citep[e.g.,][]{Sellgren:1981}. While molecular gas is in general found not to correlate with the 3.28\,$\mu$m feature in emission, strong correspondence between 3.28\,$\mu$m emission and near-infrared H$_2$ fluorescence has been seen in NGC\,2023 \citep{Gatley:1987} and Orion \citep{Sellgren:1990}, indicating an association between the emission and the presence of H$_2$ dissociation via UV photolysis. More recent studies with JWST support this picture, with the caveat that the 3.28\,$\mu$m emission is biased more toward the atomic gas than coming directly from the atomic-molecular interface like the H$_2$ \citep{Habart:2024}. The wide area 3.28\,$\mu$m emission maps from SPHEREx thus complement this picture by probing evolution of the 3.28\,$\mu$m emission and its carrier(s) from low density, diffuse gas to the transition to molecular gas, and potentially beyond.

At least some of the observed spatial offset between ice and PAH emission is attributable to geometric effects along the line of sight. As discussed in \S\ref{sec:CygX}, the CygX region contains multiple gas layers: a background layer associated with PDR interfaces (located beyond 1.4 kpc) and foreground cloud structures \citep{2024Zhang}. While the background PDR interfaces are traced by UV-heated PAH emissions \citep{Schneider2006}, the ice absorption features align with foreground dark molecular clouds, such as LDN 889, which form a layer at $\sim$950 pc \citep{2024Zhang}. As a result, PAH emission arising from the far side can be significantly attenuated by the intervening ice-rich material, and the presence of water ice absorption can further weaken its apparent intensity in the maps. This superposition not only exaggerates the apparent anti-correlation but also limits the detectability of ice in the background component, which often lacks a sufficiently bright continuum for absorption spectroscopy and is further obscured by the foreground material. Consequently, our ice maps predominantly trace the foreground molecular structures.
 
In addition to these geometrical effects, there is evidence that PAH emission is indeed responsive to local environmental conditions. In particular, regions of strong PAH emission tend to coincide with elevated dust temperatures (Spearman's rank correlation coefficient 0.834 at 36.4\arcsec resolution)(see panel~\ref{fig:Dustmaps}d and panel~\ref{fig:PAHandIce}a),
suggesting a connection with local UV irradiation \citep{Draine:2001}.

A comparison between the 3.28\,\micron\ observations and
other tracers of PAH emission is made in Figure\,\ref{fig:pah_emission}.
The IRAC data have a resolution of $\sim$2\arcsec\ and are from the Spitzer Cygnus-X Legacy project \citep{DVN/1QMXB8_2021}. The WISE data are from a custom reprocessing that removed point sources, filled in large angular scales ($>2^\circ$) with Planck data, and smoothed to a resolution of $15''$ \citep{Meisner:2014}. We employ their Tile~361 and convert to MJy\,sr$^{-1}$ by multiplying their map (in instrumental or ``DN'' units) by 0.0135 \citep{Cutri:2012}.
\begin{figure}
    \centering
    \includegraphics[width=1.00\linewidth]{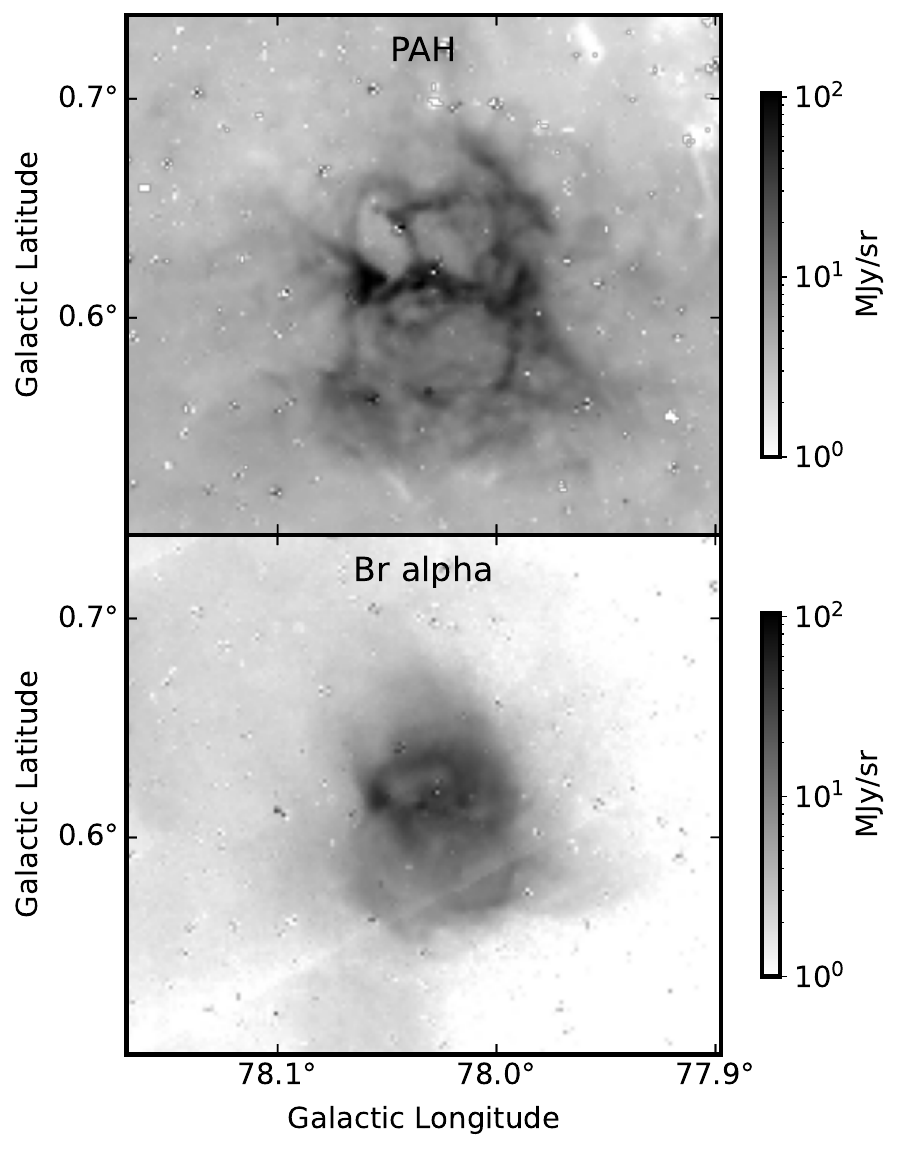}
        \includegraphics[width=1.00\linewidth]{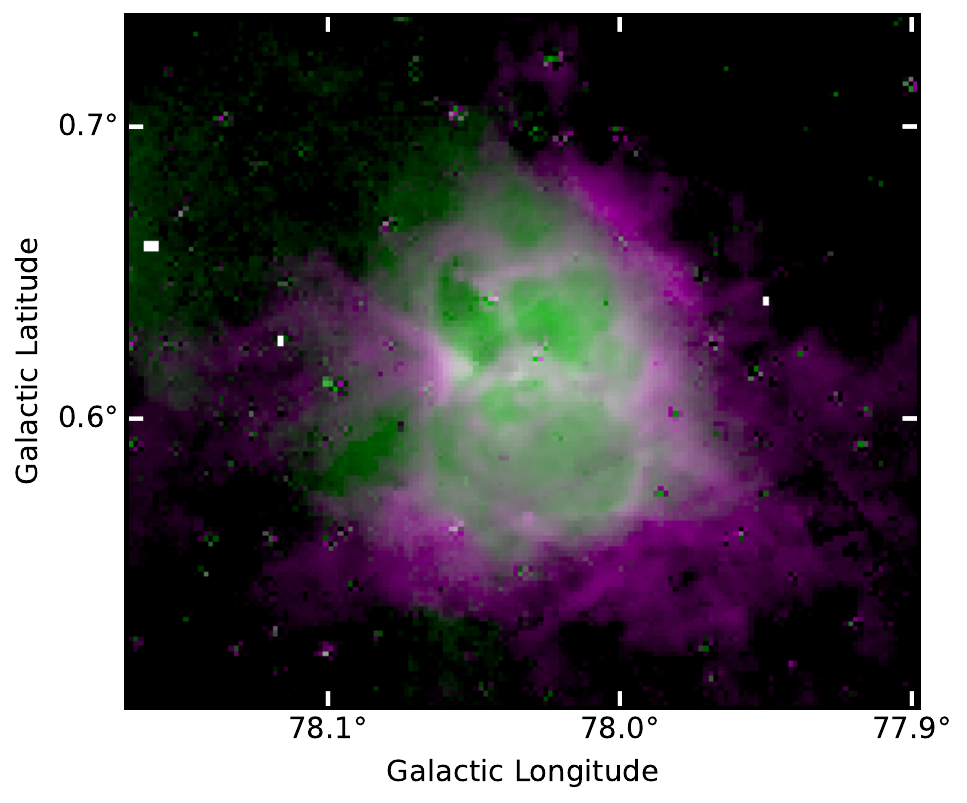}
    \caption{Images of the \ion{H}{2} region DR6 constructed from the CygX mosaics. Top: Two panels showing the continuum-subtracted 3.28\,\micron\ PAH line image and the continuum-subtracted 4.057\,\micron\ Br\,$\alpha$ line.  Bottom: a 2-color image showing the  3.28\,\micron\ PAH line image is in purple, and the Br\,$\alpha$ line is in green. The Br\,$\alpha$ emission is smoother and located in the region interior to the strongest PAH emission, which is much more filamentary.}
    \label{fig:br_alpha}
\end{figure}
While the IRAC and WISE photometric bands are broad and contain a combination of emission from PAHs, dust continuum, and various lines, the IRAC map is likely dominated by the 7.7\,\micron\ PAH feature and the WISE map by the 11.3\,\micron\ PAH feature. The strong correlation among all three maps (Spearman's rank correlation coefficient $> 0.83$ for all pairwise combinations) attests to a common source of emission for all three.

A number of astrophysical effects can affect the correlations between PAH intensities in these bands. First, as mentioned above, extinction at 3.28\,\micron\ is not negligible. Any PAH emission arising behind the cloud will undergo stronger absorption at 3.28\,$\mu$m than at the longer wavelengths. Second, as the radiation field seen by the PAHs becomes attenuated, the UV photons critical for exciting the smallest grains to emit at 3.28\,\micron\ \citep{Draine:2001, Richie:2025} become scarce. In contrast, the optical photons sufficient to produce emission in the longer wavelength features are less affected. Finally, recent evidence points to the growth of PAHs to larger sizes in moderately dense gas \citep{Zhang:2025}. This tends to weaken emission at 3.28\,\micron\ without strongly affecting the strength of the longer wavelength features \citep{Lee:2025}.

The faintest pixels in the SPHEREx 3.28\,\micron\ data in this region, with intensities less than $\sim$1\,MJy\,sr$^{-1}$, frequently correspond to regions of strong ice absorption. Here the emission in the IRAC and WISE maps is bright relative to the 3.28\,\micron\ emission, as expected. From SPHEREx 3.28\,\micron\ intensities 1--10\,MJy\,sr$^{-1}$ the relationships between the maps is remarkably linear. Above 10\,MJy\,sr$^{-1}$, comparison between the maps becomes challenging due to the differing treatments of point sources.

The maps display pronounced morphological differences in certain regions. For instance, dark clouds are visible in the 3.28\,\micron\ map where the \hto\ absorption is strong, but such features are not present in the other maps. An extended feature at ($\ell$, $b$) $\simeq (80^\circ, 0^\circ)$ is present in the WISE map but is substantially fainter in the SPHEREx and IRAC maps. Such features may be related to differences in PAH emission, but could also reflect changes in the underlying dust continuum or in the line emission that is also present in the broad IRAC and WISE bands.

Finally, Figure\,\ref{fig:pah_emission} presents the 5.0\,\micron\ SPHEREx image which shows the distribution of the warm dust continuum emission. It is highly correlated with the PAH emission maps, though distinct differences are observed. Notably, the image is less sharply filamentary than the PAH emission, with some structures completely absent. The difference in morphology is much larger than can be explained by the slightly larger PSF size at 5\,\micron. While some of the differences may be explained simply by the signal being fainter in this channel, this is unlikely to account for all of the morphological differences observed.

Disentangling the effects at play in these maps and elucidating the lifecycle of PAHs in dense gas will be the subject of future work.

\begin{figure}
    \centering
    \includegraphics[width=1.05\linewidth]{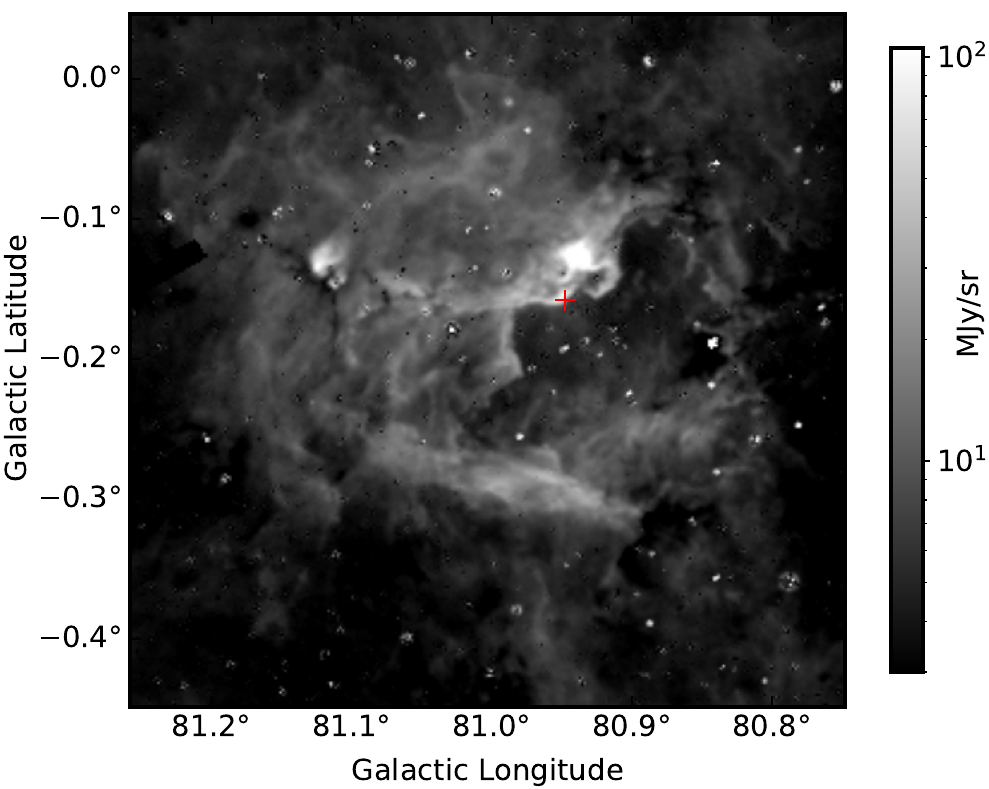}
    \vskip 5pt
    \includegraphics[width=1\linewidth]{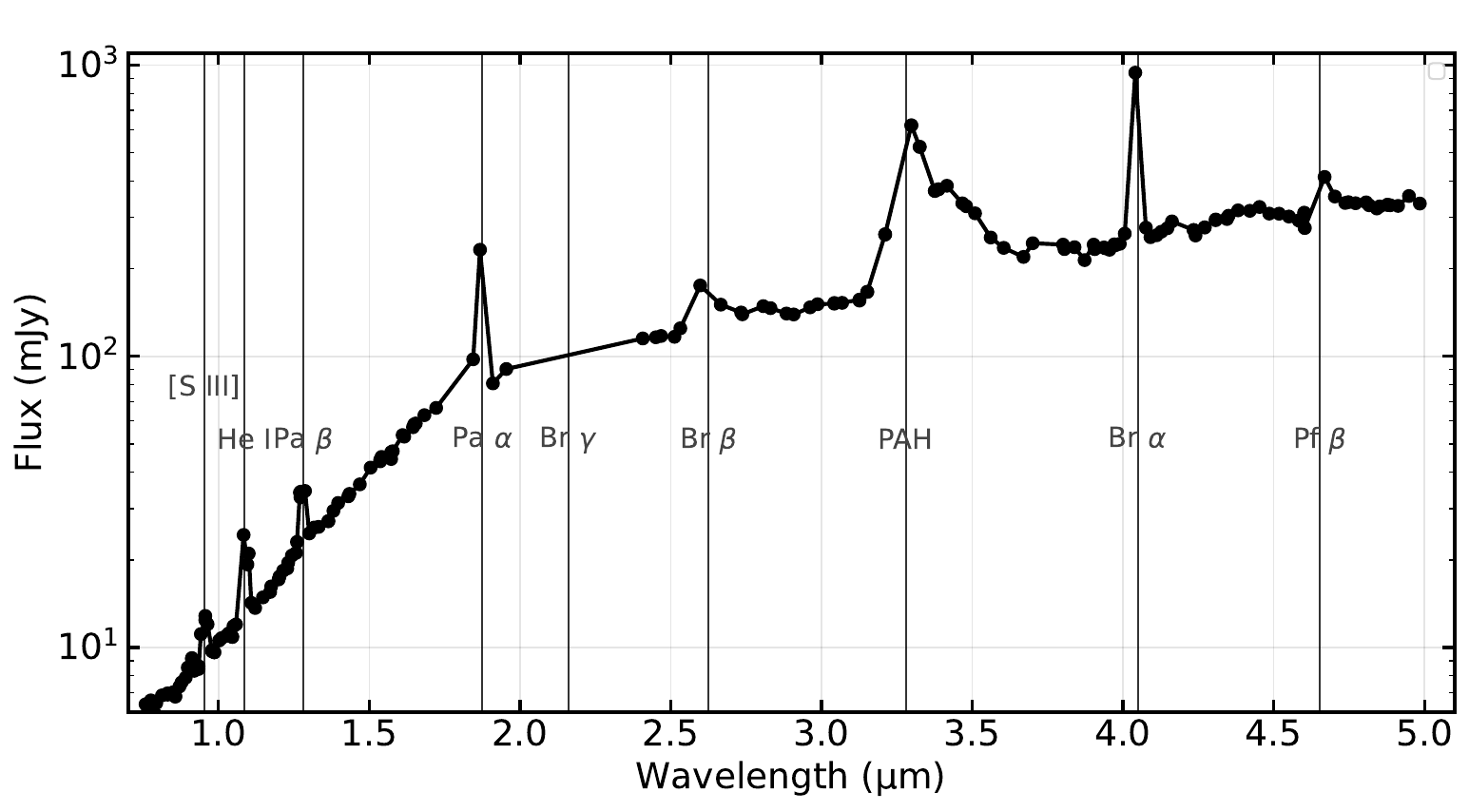}
    \caption{An extended \ion{H}{2} region (DR\,22) in CygX. The top figure shows the region sampled in the PAH line, with a red cross showing the location of the extracted spectrum. The bottom plot shows the spectrum at that location, with major emission lines labeled.}
    \label{fig:pinwheel}
\end{figure}

\subsection{Other Feature Maps and Spectra in CygX}
\subsubsection{Emission from \ion{H}{2} Regions in the Paschen, Brackett, and Pfund Series}
\label{sec:HII}

\begin{figure*}
    \centering
    \includegraphics[width=0.89\linewidth]{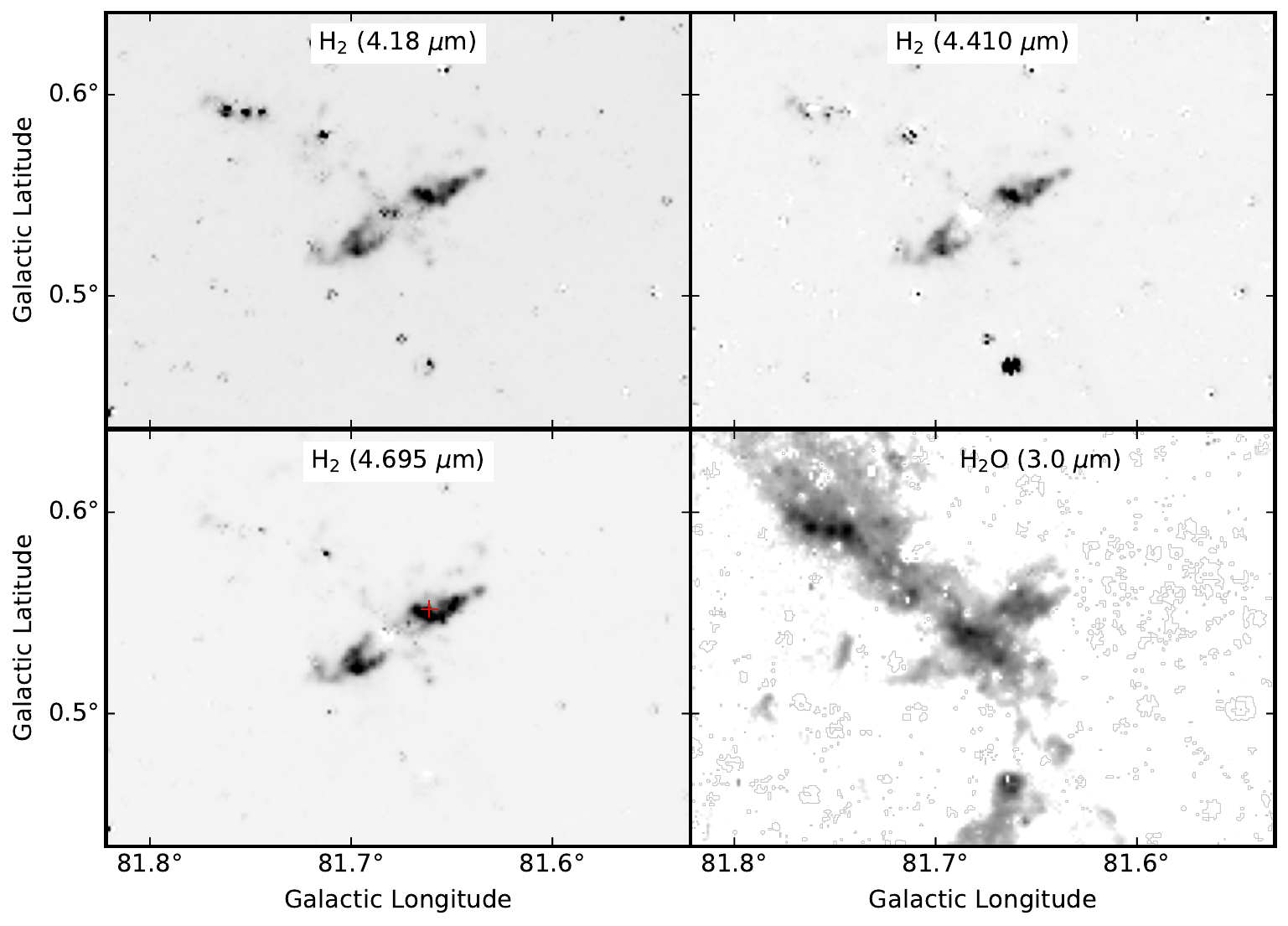}
    \caption{Continuum-subtracted line images of DR\,21, in inverse grayscale (black is higher flux) showing the strongest \h2\ lines. The top left image is centered on the 4.18\,\micron\ 0-0 S(11) line, the top right image is centered on the 4.41\,\micron\ 0-0 S(10) line, and the lower left image on the 4.695\,\micron\ 0-0 S(9) line. The lower right image shows the peak optical depth $\tau$ of the 3.05\,\micron\ \hto\ absorption feature (black is higher $\tau$). The red cross in the lower left panel marks the position of the spectrum shown in Figure\,\ref{fig:H2spec}. }
    \label{fig:h_2}
\end{figure*}
Several hydrogen recombination lines in the Paschen, Brackett, and Pfund series are accessible to SPHEREx. Brackett $\alpha$ (Br\,$\alpha$) at 4.05\,\micron\ is of particular interest because of its use as an accretion tracer and diagnostic for the conditions in YSO accretion disks \citep[e.g.,][]{1984Persson, 2013Edwards, 2020Komarova} and is at a longer wavelength than alternatives likely to be more severely absorbed by dust. For example, Br\,$\gamma$ 2.16\,\micron\ emission has been utilized as an accretion tracer in ground-based observations, but it is not as strong a line and  suffers from higher relative extinction compared to Br\,$\alpha$. The Paschen\,$\alpha$ 1.87\,\micron\ line is also typically inaccessible from the ground. SPHEREx has the capability to measure many H lines in the 0.75 -- 5\,\micron\ range, although lines between 2.42 -- 3.8\,\micron\ fall within the low-resolution ($R\sim35$) range and may be difficult to detect or extract in the presence of other lines (for example, the Brackett $\beta$ line at 2.63\,\micron).

\begin{figure}
    \centering
      \includegraphics[width=1\linewidth]{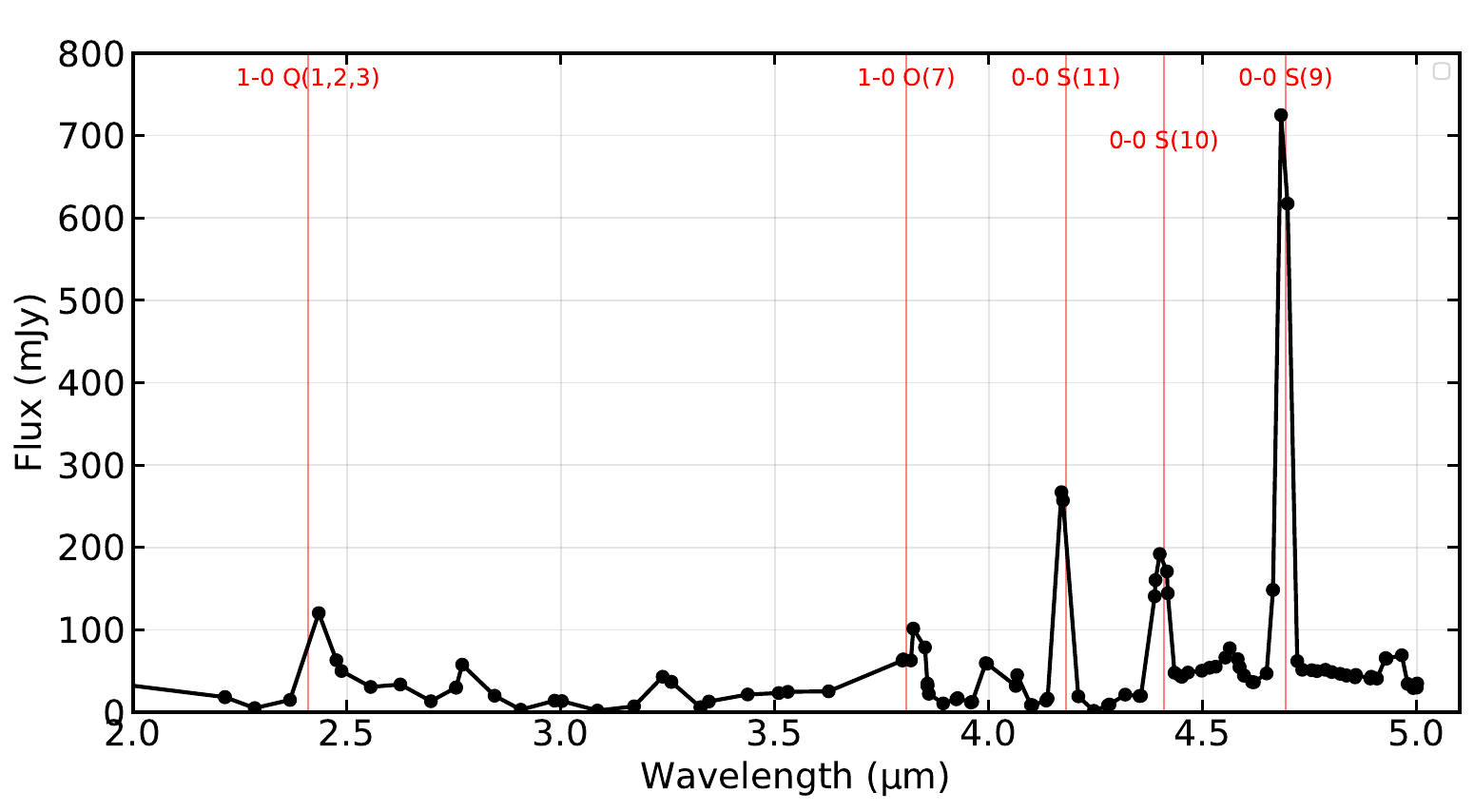}
    \caption{SPHEREx spectrum of a 18\arcsec$\times$18\arcsec\ region in the \h2 outflow of DR\,21 (centered at $l=81.6629, b=+0.5502$; marked with the cross in Figure\,\ref{fig:h_2}). The strongest \h2 lines are labeled. The line at $\sim$2.4\,\micron\ is a blend of the 1-0 Q(1), 1-0 Q(2), and 1-0 Q(3) lines.}
    \label{fig:H2spec}
\end{figure}

The lines in these series can also be used as diagnostics of conditions in \ion{H}{2} regions, and as measures of extinction as a function of wavelength. Knowledge of the extinction function can then be used to deredden other sources observed within the region \citep[][]{2023Morisset}.

An example of a region with strong Br\,$\alpha$ emission is the DR6 \ion{H}{2} region in CygX shown in Figure\,\ref{fig:br_alpha}. A narrow wavelength range (4.03 -- 4.08\,\micron) was used to construct the image. Also shown in the figure is the same region cut from the PAH image in Figure\,\ref{fig:pah_emission}. The Br\,$\alpha$ emission defines the \ion{H}{2} region that is surrounded by a more extended shell of PAH emission. This is a common structure in PDRs, where warm dust, PAHs, and other molecular material arises from 
the periphery of the \ion{H}{2} region \citep[e.g., ][]{2009Watson, 2010Watson, 2012Winston}.

Figure\,\ref{fig:pinwheel} shows another region in CygX with strong extended line emission. Aperture photometry was performed at the location indicated and the spectrum is shown in the lower panel. Lines of Pa\,$\alpha$, Pa\,$\beta$, Br\,$\alpha$, and Pf\,$\beta$ are detected on a warm dust continuum over the SPHEREx spectral range. Strong PAH emission at 3.28\,\micron\ is also seen in the spectrum. In some cases, the Pf\,$\gamma$ line at 3.297\,\micron\ can contribute to the flux measured in the PAH feature, as noted in early studies of the 3.28\,\micron\ emission \citep{1985Geballe} and is seen in a recent study of the Orion Bar \citep{2024PDRs4all}. The Pf\,$\gamma$ flux can be estimated from the Pf\,$\beta$ or Brackett series lines to determine its contribution to the 3.28\,\micron\ feature.

\subsubsection{H$_2$ Emission}
\label{sec:H2}
Emission from H$_2$ can be detected from photodissociation regions (PDRs) or from shock-excited areas such as jets or outflows. There are many diagnostic H$_2$ emission lines in the 1 -- 5\,\micron\ range that can be used to probe conditions in these regions. Ground-based studies have used the bright 2.1218\,\micron\ 1-0 S(1) line as a tracer for outflows and PDRs, but access to many other bright lines are limited because of atmospheric absorption and high backgrounds $> 2.5$\,\micron.  

\spherex\ makes it possible to access the \h2\ lines at wavelengths not readily available from the ground. As an example, the region around the massive protostar DR\,21 is shown in Figure\,\ref{fig:h_2}. Its H$_2$ outflow has been studied extensively \citep[e.g.,][]{1986Garden, 1990Garden,1991Garden,1996Davis,1997Fernandes,1998Smith}.  \citet{2006Smith} examined the IRAC images and determined that a large fraction of the emission detected in the 4.5\,\micron\ band was due to shocked H$_2$ emission.

The SPHEREx images in Figure\,\ref{fig:h_2} and the spectrum in Figure\,\ref{fig:H2spec}  demonstrate that the jet is emitting strongly in H$_2$. The SPHEREx spectrum extracted from a 18\arcsec$\times$18\arcsec\ region in the outflow of DR\,21 confirms that the infrared emission is dominated by H$_2$ lines. Using the two brightest lines, we estimate an excitation temperature of the H$_2$ gas of 2300\,K, consistent with collisionally heated gas. The images of DR\,21 in the \hto, CO$_2$, and CO lines show ice absorption along the ridge perpendicular to the outflow and no emission in the outflow, confirming that the emission detected in the IRAC images is primarily H$_2$ emission from the shock region. The lower right panel in Figure\,\ref{fig:h_2} shows the peak optical depth of the 3.05\,\micron\ \hto\ absorption line.

\section{Conclusions}
\label{sec:conclusions}
The 5\degr$\times$5\degr\ maps we have presented of the CygX region demonstrate SPHEREx's ability to produce wide-field maps of spectral features in the 0.75 -- 5\,\micron\ wavelength range. The maps of \hto\ and \co2\ ice absorption show the distribution of these species in filamentary structures that extend for many degrees across the field. The distribution of these ices is similar, however the spectra of background sources behind the dark clouds show the relative strengths (and therefore column densities) of the absorption features vary on small spatial scales. This effect is also seen in the many lines of sight probed in the LDN\,935 dark cloud.

A comparison of the \hto\ and \co2\ ice maps with NICEST $J$-band extinction, $^{13}$CO emission, and Herschel-derived dust column density reveals a consistent physical picture across CygX. Ice absorption is strongest along cold, dense, and well-shielded lines of sight, broadly following regions of high column density and enhanced $^{13}$CO emission while avoiding warmer environments. The remaining differences likely arise from gas-phase CO freeze-out in cold, dense regions and from line-of-sight geometric effects that limit the contribution of more diffuse or background cloud components to the ice absorption maps, even though these components are still traced by other dust and gas diagnostics. Overall, these results are fully consistent with the established picture of ice formation in dense molecular clouds and demonstrate that this picture remains valid over the large spatial scales probed by CygX, encompassing multiple giant molecular clouds.

The SPHEREx 3.28\,\micron\ PAH image is strongly correlated with the IRAC and WISE images that sample the 7.7 and 11.3\,\micron\ features, but differences are seen which may be related to grain size effects and/or the UV radiation field. The higher spectral resolution of SPHEREx compared to the broad bandpasses used by Spitzer/IRAC and WISE enable separating the PAH feature from continuum and other line emission in the 3-4\,\micron\ wavelength range, which will allow for accurate measurements of the PAH feature, as well as the study of aliphatic grains in these regions. Comparison of the PAH emission to the ice distribution shows an anti-correlation which can be explained by the PAH emission being strongest in regions of higher dust temperature, whereas the ice absorption is higher in dense regions that are shielded from UV radiation.

We have shown that in addition to ice absorption features and PAH emission, SPHEREx detects hydrogen recombination lines and \h2\ from many regions which are sampled in its 0.75 -- 5\,\micron\ spectral range. We have presented several images and spectra showing how SPHEREx can be used to study the emission from these species in point sources and in extended emission regions of the ISM.

As this paper is submitted, SPHEREx has completed one full map of the sky. During the remainder of the nominal two year mission, SPHEREx will continue to observe and map the full sky three more times. The SPHEREx Science Team has produced all-sky images from the first surveys\footnote{\url{https://www.jpl.nasa.gov/images/pia26600-spherexs-first-all-sky-map/}} \citep{2026Cukierman}, and all-sky datacubes in 102 spectral channels are planned to be released 6 months after the first year of the nominal mission \citep[][see the data release schedule\footnote{\url{https://spherex.caltech.edu/page/data-products}}]{2025Akeson}. 

In addition to the main mission themes, the SPHEREx maps and 0.75 - 5.0\,\micron\ spectra will support many additional science investigations in the Milky Way and nearby galaxies.  Numerous lines of astrophysical interest reside in this wavelength range, including emission from hydrogen recombination lines in the Paschen, Brackett, and Pfund series, molecular hydrogen (\h2), polycyclic aromatic hydrocarbons (PAHs), the CO fundamental v = 1–0 ro-vibrational band around 4.7\,\micron\ and overtone 2-0 band around 2.3\,\micron, and thermal emission from warm dust. 

\begin{acknowledgements}

We acknowledge support from the \spherex\ project under a contract from the NASA/Goddard Space Flight Center to the California Institute of Technology.

Part of the research described in this paper was carried out at the Jet Propulsion Laboratory, California Institute of Technology, under a contract with the National Aeronautics and Space Administration (80NM0018D0004).

This work is based in part on observations made with the NASA/ESA/CSA James Webb Space Telescope. The data were obtained from the Mikulski Archive for Space Telescopes at the Space Telescope Science Institute, which is operated by the Association of Universities for Research in Astronomy, Inc., under NASA contract NAS 5-03127 for JWST. 

We acknowledge the use of the Smithsonian
High Performance Cluster (SI/HPC), Smithsonian Institution,
DOI: \href{https://doi.org/10.25572/SIHPC}{10.25572/SIHPC}
The authors acknowledge the Texas Advanced Computing Center (TACC)\footnote[2]{\url{http://www.tacc.utexas.edu}} at The University of Texas at Austin for providing computational resources that have contributed to the research results reported within this paper.
J.-E. Lee and J. K. Noh were supported by the National Research Foundation of Korea (NRF) grant funded by the Korea government (MSIT; grant Nos. 2021R1A2C1011718 and RS-2024-00416859).

\end{acknowledgements}

\bibliography{main}{}
\bibliographystyle{aasjournalv7}

\end{document}